\documentclass[12pt]{iopart}

\usepackage{graphicx}
\usepackage{mathptmx, courier, pifont}
\usepackage[scaled=0.92]{helvet}
\usepackage[T1]{fontenc}
\usepackage{textcomp}

\newlength{\figdn}
\newcommand{\fplus}[1]{F^+_{#1}}
\newcommand{\fminus}[1]{F^-_{#1}}

\newcommand{\eq}[1]{Eq.~(\ref{#1})}
\newcommand{\eqnoeq}[1]{(\ref{#1})}
\newcommand{\fig}[1]{Fig.~\ref{fig:#1}}
\newcommand{\diffelement}[1]{d#1}
\newcommand{\deriv}[2]{\frac{d#1}{d#2}}

\newcommand{\tsub}[1]{_{\mbox{\scriptsize#1}}}
\newcommand{\tsup}[1]{^{\mbox{\scriptsize#1}}}
\newcommand{\tfrac}[2]{\mbox{$\small\frac{#1}{#2}$}}
\newcommand{\units}[1]{\mbox{\ #1}}
\newcommand{\isotope}[2]{\mbox{$^{#1}$#2}}

\newcommand{\singlefig}[6]{%
\begin{figure} \vspace{#3}%
\begin{flushright}%
\includegraphics*[scale=#5]{#2}%
\end{flushright}%
\caption{\label{fig:#1} #6}%
\vspace{#4}%
\end{figure}}

\newcommand{\putfig}[6]{%
\begin{figure}\vspace*{#2}%
\begin{flushright}%
\includegraphics*[scale=#4]{#5}%
\end{flushright}%
\caption{#6}%
\vspace*{#3}%
\label{fig:#1}%
\end{figure}}

\begin{document}

\title{Explicit Integration of Extremely-Stiff Reaction Networks: Asymptotic
Methods}

\author{
M. W. Guidry$^{1,2,3}$, 
R. Budiardja$^{1,2,3}$,
E. Feger$^{1}$,
J. J. Billings$^{3}$,
W. R. Hix$^{1,2,3}$,
O. E. B. Messer$^{4,1}$,
K. J. Roche$^{5,6}$,
E. McMahon$^{1}$,
and M. He$^{7,1}$
}

\address{$^1$Department of Physics and Astronomy, University of Tennessee,
Knoxville, TN 37996-1200, USA}
\address{$^2$Physics Division, Oak Ridge National Laboratory, Oak Ridge, TN
37830, USA}
\address{$^3$Computer Science and Mathematics Division, Oak Ridge National
Laboratory, Oak Ridge, TN 37830, USA}
\address{$^4$National Center for Computational Sciences, 
Oak Ridge National Laboratory,  Oak Ridge, TN 37831, USA}
\address{$^5$Pacific Northwest National Laboratory, Richland, WA 99352,
USA.}
\address{$^6$Department of Physics, University of Washington, Seattle,
WA 98195, USA.}
\address{$^7$Department of Mathematics, Shanghai Jiao-Tong University,
Shanghai, PRC}

\ead{guidry@utk.edu}

\begin{abstract} We show that, even for extremely stiff systems, explicit
integration  may compete in both accuracy and speed with  implicit methods if
algebraic methods are used to stabilize the numerical integration. The required
stabilizing algebra depends on whether the system is well-removed from
equilibrium or near equilibrium. This paper introduces a quantitative
distinction between these two regimes and addresses the former case in depth,
presenting explicit asymptotic methods appropriate when the system is extremely
stiff but only weakly equilibrated. A second paper \cite{guidQSS} examines
quasi-steady-state methods as an alternative to asymptotic methods in systems
well away from equilibrium and a third paper \cite{guidPE} extends these methods
to equilibrium conditions in extremely stiff systems using partial equilibrium
methods. All three papers present systematic evidence for timesteps competitive
with implicit methods. Because an explicit method can execute a timestep faster
than an implicit method, algebraically-stabilized explicit algorithms might
permit integration of larger networks than have been feasible before in various
disciplines. \end{abstract}

%Uncomment for PACS numbers title message

% Keywords required only for MST, PB, PMB, PM, JOA, JOB? 
\vspace{2pc}

% Uncomment for Submitted to journal title message
%\submitto{\JPA}

% Comment out if separate title page not required
%\maketitle

\section{Introduction}

In many scientific and technical contexts one encounters phenomena that may be
modeled by fluxes transferring population between sources and sinks for various
species.  Examples include kinetic processes that modify abundances and transfer
energy in atomic, molecular, and nuclear systems;  geochemical, climate, and
other environmental systems; electrical circuits; economic models; and
population dynamics. Terminology varies  but let us refer generically to these
sources and sinks as {\em boxes,} and term the resulting systems of boxes
connected by fluxes {\em reaction networks.} Such systems are commonly modeled
by a coupled set of differential equations that describe a continuous flow of
population through the boxes.  

The reaction network is often classified as a {\em stiff system}, which we shall
define to be a system of equations containing multiple timescales ranging over
many orders of magnitude \cite{gear71,lamb91,press92,oran05}. Most  physical
systems involve important processes operating on very different timescales, so
realistic problems tend to be at least moderately stiff. Some, such as those
encountered in many astrophysics applications, are extremely stiff, with fastest
and slowest timescales in the problem differing by as much as 10--20 orders of
magnitude. In stiff systems the timestep constraints are set by numerical
stability requirements rather than accuracy considerations. Hence, explicit
numerical integration of stiff systems is usually impractical because the
maximum stable timestep is far too small for efficient solutions (see, for
example, Refs.\ \cite{press92,oran05}).  This is commonly addressed by employing
implicit or semi-implicit stiff solvers that are stable, but that require
time-consuming iterative matrix solutions.  

A given box in a reaction network often is connected strongly only to a few
other boxes.  For example, the explosive burning conditions encountered in
astrophysical novae, X-ray bursts, or supernovae may require reaction networks
with hundreds to thousands of nuclear isotopes.  Yet individual isotopes are
typically connected directly to other isotopes through (at most) $\sim$10
reactions of consequence, and under many conditions no more than 2--3 reactions
are important for a given isotope. Such restrictions on the direct box reaction
coupling imply that the matrices appearing in the iterative implicit solution
are {\em sparse.} Although various methods are available to deal with sparse
matrices, in practice many codes for solving large reaction networks have not
exploited sparseness in particularly effective ways.  

For example, in astrophysical calculations with implicit solvers in large
networks (say $\sim 150$ species or more), one finds often that greater than
90\% of the processor time is consumed in matrix operations \cite{timmes,hix05}.
Efficient algorithms exist for the required matrix algebra (with incremental
improvements in them over time), but the matrix nature of the core problem
implies that the time required for implicit solution grows non-linearly with the
size of the network. In typical working codes for large-scale applications,
increasing the size of the network increases the time for solution, often
quadratically, sometimes as much as cubically, until there are enough boxes in
the network to justify the overhead of sparse-matrix methods with more favorable
scaling. In applications in thermonuclear networks, for example, it is often
found that the overhead required to implement sparse-matrix iterative solutions
is not justified until there are several hundred boxes in the network. Thus,
many present implicit stiff-network algorithms do not scale very gracefully to
larger networks.

We are primarily interested in the most ambitious applications of large
networks, where the reaction network is only a portion of a larger problem.  Let
us take as representative astrophysical thermonuclear reaction networks, where a
proper description of the overall problem typically requires multi-dimensional
hydrodynamics or radiation hydrodynamics coupled tightly to a large
thermonuclear reaction network.  The hydrodynamical evolution controls the
conditions in the network such as temperature and density, and the network
influences the hydrodynamic evolution strongly through energy production and
modification of composition variables. As a consequence of the limitations
discussed in the preceding paragraphs, the solution of large networks by the
usual approaches is time-consuming and few calculations have attempted to couple
the element and energy production strongly to the hydrodynamics with a network
of realistic complexity.  The most ambitious approaches use very small networks,
perhaps tuned empirically to get critical quantities like energy production
correct on average, coupled to the hydrodynamical simulation.  In many
calculations even this is not done and the network is replaced entirely by
parameterization.  Then a more complete network is run in a separate
``post-processing'' step, where fixed hydrodynamical profiles computed in the
hydrodynamical simulation with the small network are used to specify the
variation of thermodynamic variables such as temperature and density with time.

Astrophysical thermonuclear networks have been used for illustration, but many
problems of scientific and technical interest  exhibit similar complexity. 
Examples include astrochemical kinetics, where one must model large chemical
evolution networks in contracting molecular clouds, or combustion chemistry,
where chemical burning networks are strongly coupled to simulations of the
dynamics of the air and fuel mixture. Physically-realistic networks in such
contexts would often be quite large. In combustion of larger hydrocarbon
molecules or studies of soot formation, hundreds to thousands of reacting
species undergoing as many as 10,000 reactions may be encountered \cite{oran05},
and in supernova explosions hundreds to thousands of nuclear isotopes with tens
of thousands of reaction couplings make non-zero contributions \cite{hix05}. For
such cases one finds that current techniques do not allow for a coupling of
realistic reaction networks to the full dynamics of the problem and often
severely truncated or highly schematic reaction networks have been used in even
the most realistic simulations.

% See Table 5.6 in Oran and Boris for above

\section{Reaction Networks in the Context of Larger Problems}

To be definite, we shall assume that the coupling of reaction networks is done
using operator splitting, where the hydrodynamical solver  is evolved for a
numerical timestep holding network parameters constant, and then the network is
evolved over the time corresponding to the hydrodynamical timestep holding the
new hydrodynamical variables constant.  This places two basic constraints on
methods:

\begin{enumerate}

 \item 
At the end of each hydrodynamical timestep the network must be advanced with new
initial conditions.  Thus, algorithms must be capable of rapid initialization
and must not depend in a complex way on conditions from previous time
intervals. 

\item
With modern processors, existing algorithms are reasonably adequate for many
post-processing calculations. In contrast, for the operator-split, parallel
processing environment that is our interest here, solution of the network over a
hydrodynamic timestep must be fast enough that it does not require time
substantially larger than that for the hydrodynamical solution.  

\end{enumerate}
Let us elaborate further on this second point. If a single processor were used
to calculate both the hydrodynamical evolution and the network evolution in one
hydrodynamical zone, the network evolution over a hydrodynamical timestep
interval must be fast enough to not slow the calculation by too much relative to
the hydrodynamical evolution alone. If we take the point of view that we are
willing to tolerate longer compute times in the interest of a much more
realistic calculation, but not longer by orders of magnitude, we estimate that
the network must be capable of evolving over the time interval corresponding to
the hydrodynamical timestep in roughly a second or less wall clock time. 

We take the multidimensional, adaptive-mesh, explicit hydrodynamical FLASH code
[9] applied to Type Ia supernova simulations on large parallel systems as
representative. The explicit hydrodynamical timestep will be limited overall by
the Courant time (roughly, because stability requires a hydrodynamical timestep
not larger than the sound-crossing time for the zone), and more stringently in
zones of rapid burning where temperature and density may be changing rapidly. In
current Type Ia supernova simulations the Courant time would typically be
$10^{-4}$ s or smaller over most of the grid for the timescale relevant for the
main part of the explosion, with rapid nuclear burning and associated
temperature changes limiting the hydrodynamical timestep to
$10^{-8}$--$10^{-10}$ s for some ranges of times. For qualitative estimates, let
us take as representative that a typical network integration for a single
hydrodynamical timestep will be over an interval $\sim 10^{-8}$ s during the
time of strong burning and $\sim 10^{-6}$ s over much of the approach to
equilibration after strong burning.

In FLASH,  many spatial  zones will be assigned to a single MPI rank on a
parallel system. Therefore, in the absence of node-level parallelism (for
example, with OpenMP), the network must be capable of calculating a number of
hydrodynamic time intervals in a second or less if we wish to calculate an
independent network for each zone. Let us take for estimation purposed that we
wish
to be able to reliably integrate 1000 independent networks over a time interval
of say $10^{-6}$ seconds on a single processor in one second wall clock time,
with each network containing several hundred isotopes. This places extremely
strong startup and speed constraints on the required network. The explicit
algorithms discussed here are capable of  perhaps $10^4$ network timesteps per
second on a single  processor with present technology for a network with $\sim
150$ isotopes, so our goals require a network algorithm that can integrate a
time interval of order $10^{-6}$ seconds in no more than $\sim 10-100$
timesteps, implying average stable and accurate timesteps at least as large as
$10^{-2} - 10^{-3}$ times the elapsed integration time for the corresponding
hydrodynamical evolution. Figure \ref{fig:dtHydro} illustrates.

\putfig
{dtHydro}
{0pt}
{\figdn}
{0.5}
{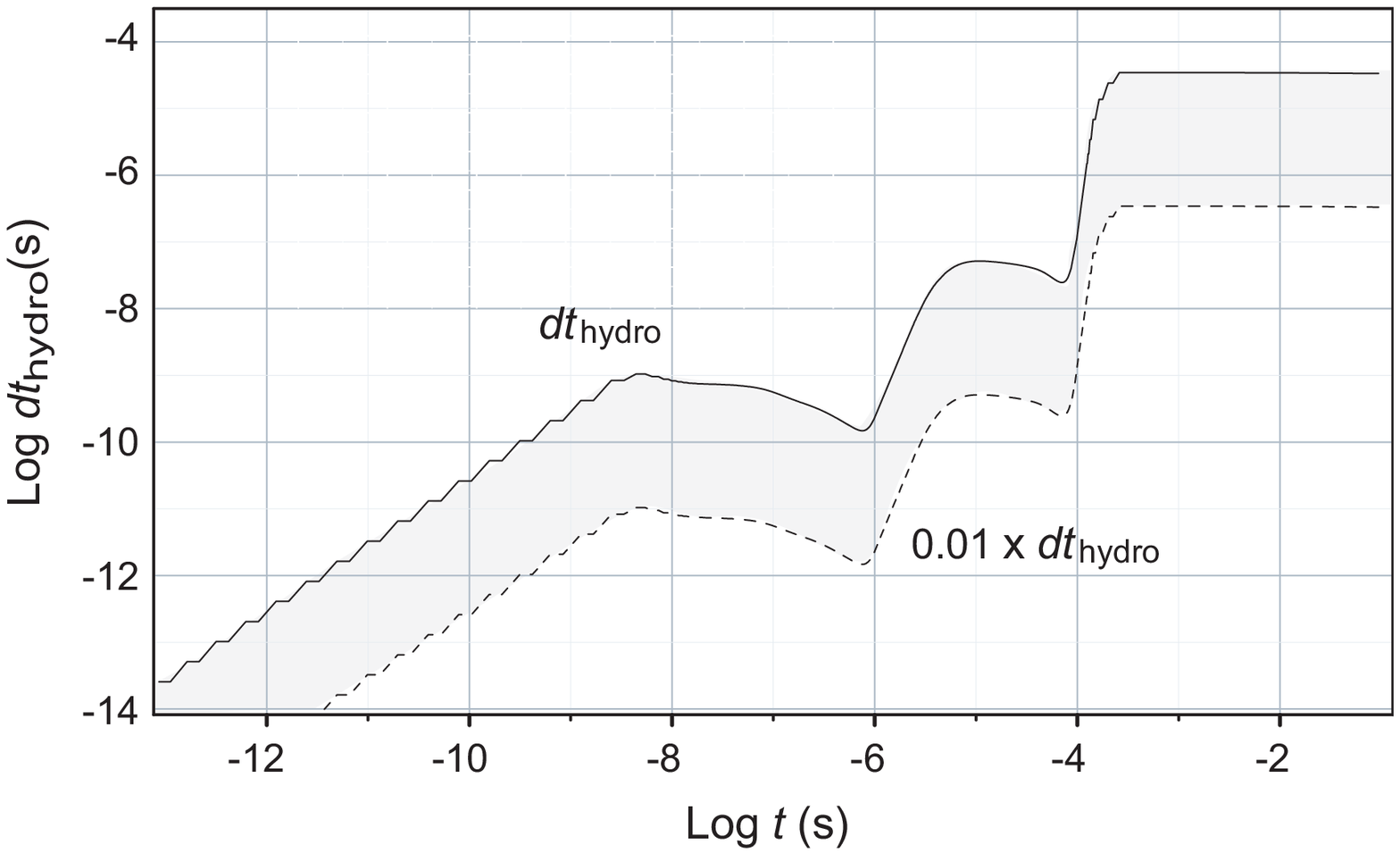}
{Hydrodynamical timesteps in a typical Type Ia supernova simulation
(solid upper curve). The calculation corresponds to a single zone integrated
with the Flash hydrodynamical code \cite{fry00}, operator-split coupled to a
150-isotope network using the explicit asymptotic method described below.
Initial conditions were equal mass fractions of \isotope{12}{C} and
\isotope{16}{O}, with an initial temperature of $3 \times 10^9$ K and initial
mass density of $10^7 \units{g\,cm}^{-3}$. The shaded region between the two
curves represents the range of timesteps an explicit network calculation must
take to be able to integrate 100-1000 zones on a single processor over one
operator-split hydrodynamical timestep in less than about one second elapsed
time on modern processors. The hydrodynamical timestep lies roughly in the range
$0.1 t$--$0.001 t$, where $t$ is the elapsed time, over most of the range of
integration. To maintain network timesteps within the band for each
operator-split hydrodynamical timestep, we see that the algorithm must be
capable of taking stable and accurate network timesteps approximately in the
range $0.01t$--$0.0001 t$ over the entire range of hydrodynamical integration.}

Such large timesteps are often possible with implicit and semi-implicit
algorithms, but those methods are inefficient at computing each timestep;
explicit methods can compute a timestep efficiently, but timesteps this large
are unthinkable with a normal explicit algorithm because they would be unstable
in most realistic situations.   In this and the other two papers
\cite{guidQSS,guidPE} of this series we shall demonstrate stabilization methods
for explicit integration that realize such competitive integration timesteps in
a variety of examples. Thus we shall reopen the discussion of whether explicit
methods, with their faster computation of timesteps and more favorable scaling
with network size, are practical for large, stiff networks.

\section{\label{sh:large-stiff} Stiffness in Reaction Networks}

The general task is to solve efficiently $N$ coupled ordinary differential
equations 
\begin{eqnarray}
    \frac{dy_i}{dt} &=& F_i(y,t) = \sum_j F_{ij}(t) 
\nonumber
\\
&\equiv& \fplus i (t)- \fminus i(t)
    = \fplus i (t) - k_i(t) y_i(t)
\label{eq1.1}
\end{eqnarray}
subject to appropriate boundary conditions.  In this expression, the $y_i (i=1
\dots N)$ describe the dependent variables (typically measures of abundance),
$t$ is the independent variable (the time in our examples), the fluxes between
species $i$ and $j$ are denoted by $F_{ij}$, and $k_i(t)$ is the effective rate
for all processes depleting the species $i$.  The sum for each variable $i$ is
over all species $j$ coupled to $i$ by a non-zero flux $F_{ij}$, and for later
convenience we have decomposed the flux into a component $\fplus i$ that
increases the abundance of $y_i$ and a component $\fminus i = k_i y_i$ that
depletes it. For an $N$-species network there will be $N$ such equations in the
population variables $y_i$, generally coupled to each other because of the
dependence of the fluxes on the different $y_j$. (For notational simplicity we
will not always display the $i$ index explicitly on the right side in our
equations). The variables $y_i$ are typically proportional to a number density
for the species $i$. To keep the discussion general the variable $y_i(t)$ will
be used in most of our equations, but for the specific astrophysical examples
that follow we shall replace the generic population variables $y_i$ with the
mass fraction $X_i$, which satisfies
\begin{equation}
X_i  = \frac{n_iA_i}{\rho N\tsub A}
\qquad
\sum_i X_i =1,
\label{5.35}
\end{equation}
where $N\tsub A$ is Avogadro's number, $\rho$ is the total mass density, and
$A_i$ is the atomic mass number and $n_i$ the number density for the species
$i$.

\subsection{\label{sh:cno} Example: Stiffness and Stability in the CNO Cycle}

The carbon--nitrogen--oxygen (CNO) cycle that powers main-sequence stars more
massive than the Sun provides a graphic illustration of stiffness in a
relatively simple system of large physical significance. The CNO cycle is
displayed in Fig.~\ref{fig:cnoCycle}. 
\putfig
{cnoCycle}
{0pt}
{\figdn}
{0.90}
{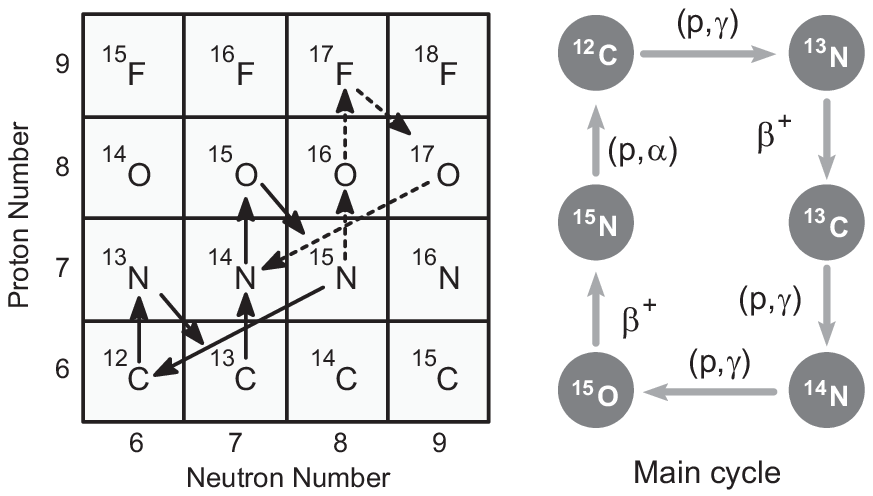}
{The CNO cycle. On the left side the main branch of the cycle is illustrated
with solid arrows and a side branch is illustrated with dashed arrows. On the
right side, the main branch of the CNO cycle is illustrated in more detail.}
We shall illustrate by considering the primary part of the CNO cycle illustrated
on the right side of the figure. If the thermonuclear network corresponding to
the main part of the CNO cycle is integrated under typical CNO cycle temperature
and density conditions by explicit forward Euler methods using standard rates
\cite{raus2000} for the reactions and constant timesteps, the integration is
stable for timesteps less than or equal to 285.7 seconds, but becomes
catastrophically unstable for a timestep of 285.8 seconds or more. This
instability threshold is {\em precisely two over the fastest rate} for the
transitions in the network, which corresponds to the $\beta$-decay of
\isotope{15}O to produce \isotope{15}N.

We now show that this instability for forward differencing of the CNO cycles
arises because rapidly-decreasing small populations can become negative in an
explicit integration if the timestep is too large.  These negative populations
export unphysical negative population that can destabilize the system because
they can lead to exponentially growing solutions in small components that
ultimately couple to the larger components.  Let us elaborate through the use of
a simple model illustrated in Figs.\
\ref{fig:negProb}--\ref{fig:O15LogAbsoluteDiverge}, which will generalize a
discussion that may be found in Ref.\ \cite{oran05}.
\putfig
{negProb}
{0pt}
{\figdn}
{0.65}
{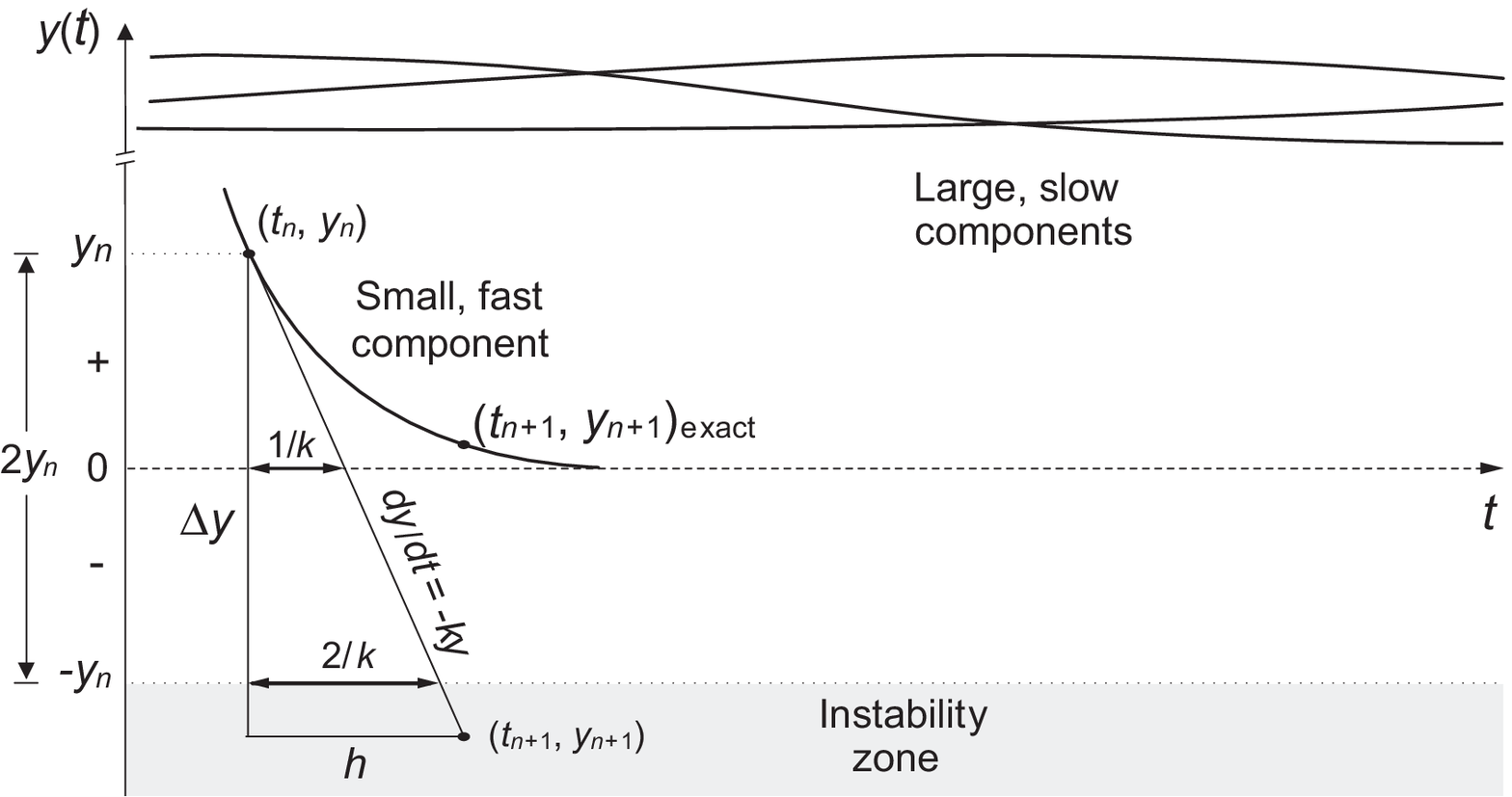}
{One common origin of stiffness instability.  For simplicity in discussion the
timescale $h$ is presumed to be sufficiently short that the large, slow
components remain essentially unchanged over a time equal to many times $h$.
The explicit integration for the small, fast component diverges if  $y_{n+1} \le
-y_n$, such that $y_{n+1}$ lies in the shaded region.  This divergence is
exponential for longer times and quickly propagates to the larger slow
components, destabilizing the entire network.  The illustration of the slow
components is schematic; in typical stiff systems their timescales may be many
orders of magnitude longer than those for fast components and often they are
larger in value by many orders of magnitude.}

\subsection{\label{ss:simpleMod}A Simple Model for One Form of Stiffness}

In Fig.\ \ref{fig:negProb} we assume a coupled system in which there are  many
components varying on a relatively long timescale (illustrated schematically by
the curves at the top) and a single component that is small and exponentially
decreasing on a much faster timescale, with a typical behavior $y(t) =
\exp(-kt)$ so that $dy(t)/dt = -ky(t)$.  Let us consider timesteps that remain
small compared with the larger timescales in the system (so that for a single
step we are in an approximately adiabatic situation and the populations varying
on long timescales may be considered frozen), but comparable to or larger than
the timescale set by $1/k$.

To advance the solution for the fast component from $t_n$ to $t_{n+1}$ by the
explicit Euler method, we take a timestep $h$ and extrapolate the solution using
the derivative $dy/dt = -ky$ evaluated at $(t_n,y_n)$.  The triangle of height
$\Delta y$ in Fig.\ \ref{fig:negProb} summarizes, where $(t_{n+1},y_{n+1})$
represents the numerical solution and $(t_{n+1},y_{n+1} )\tsub{exact}$ is the
exact solution. From this construction we see that \cite{oran05}

\begin{enumerate}
\item
For $h<1/k$, the forward Euler approximant for step $n+1$ will yield a value of
$y_{n+1}$ that lies between 0 and $y_n$, so that $|y_{n+1}| < |y_n|$.

\item
For $1/k < h <2/k$, the sign of $y_{n+1}$ will be negative but again $|y_{n+1}|
< |y_n|$.

\item
For $h\ge 2/k$ the sign of $y_{n+1}$ will be negative and $|y_{n+1}| \ge |y_n|$.

\end{enumerate}
But for the forward Euler approximation to $y(t) = \exp(-kt)$ we have $y_{n+1} =
(1-hk)y_n$, so that by iterating for $m$ successive steps of fixed size $h$
(with $mh$ still considerably less than the longer timescales in the system, so
that the adiabatic approximation remains valid), we obtain
\begin{equation}
   y_{n+m} = (1-hk)^m y_n.
\label{negative1.2}
\end{equation}
This converges toward the correct value of zero at larger times only if the
product $hk$ lies between zero and two, implying that the maximum value of $h$
that yields a convergent solution is bounded by $h < 2/k$.  Thus, in Fig.\
\ref{fig:negProb} the maximum stable value of $h$ is less than $2/k$ and any
point $(t_{n+1},y_{n+1})$ extrapolated by forward Euler integration that lies in
the shaded zone (that is, $y_{n+1}  \le -y_n$) will be unstable under the
iteration (\ref{negative1.2}), diverging to infinity rather than converging to
the correct value of zero.  Because the small, fast components are coupled to
the other components of the network, this divergence will quickly destabilize
the entire network. Let us illustrate with a concrete example.  Figure
\ref{fig:compositeConvergenceAbridged}%
\putfig
{compositeConvergenceAbridged}
{0pt}
{\figdn}
{0.88}
{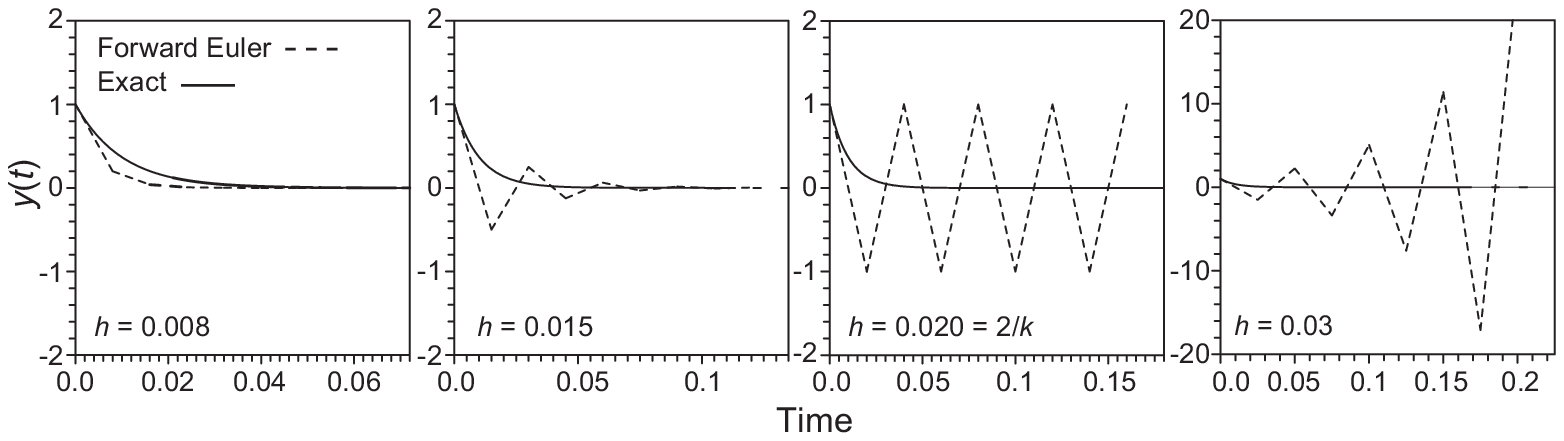}
{Behavior of  $\exp(-kt)$ with $k=100$ for increasing explicit timesteps
$h$. 
Generally for $ h < 2/k$ the solution converges to the correct value of zero but
for $h>2/k$ the solution diverges to $\pm \infty$ under successive iterations.
Note that the vertical scale has been increased by a factor of 10 in the
rightmost figure.}
 corresponds to forward Euler solution of $y_{n+1} = (1-hk)y_n$ with $k=100$ 
and timesteps ranging between $h=0.008$ and 0.03.
In this example, we see that 

\begin{enumerate}

\item
For $ h < 2/k$, the solution 
converges to zero.  

\item
For $h=2/k$, the solution oscillates between positive and 
negative values of the same absolute value and neither converges nor diverges.  

\item
For $h>2/k$, the solution diverges to $\pm \infty$ under 
successive iterations, with the divergence exhibiting exponential behavior for
larger times.

\end{enumerate}

Figure \ref{fig:O15LogAbsoluteDiverge}%
\putfig
{O15LogAbsoluteDiverge}
{0pt}
{\figdn}
{0.92}
{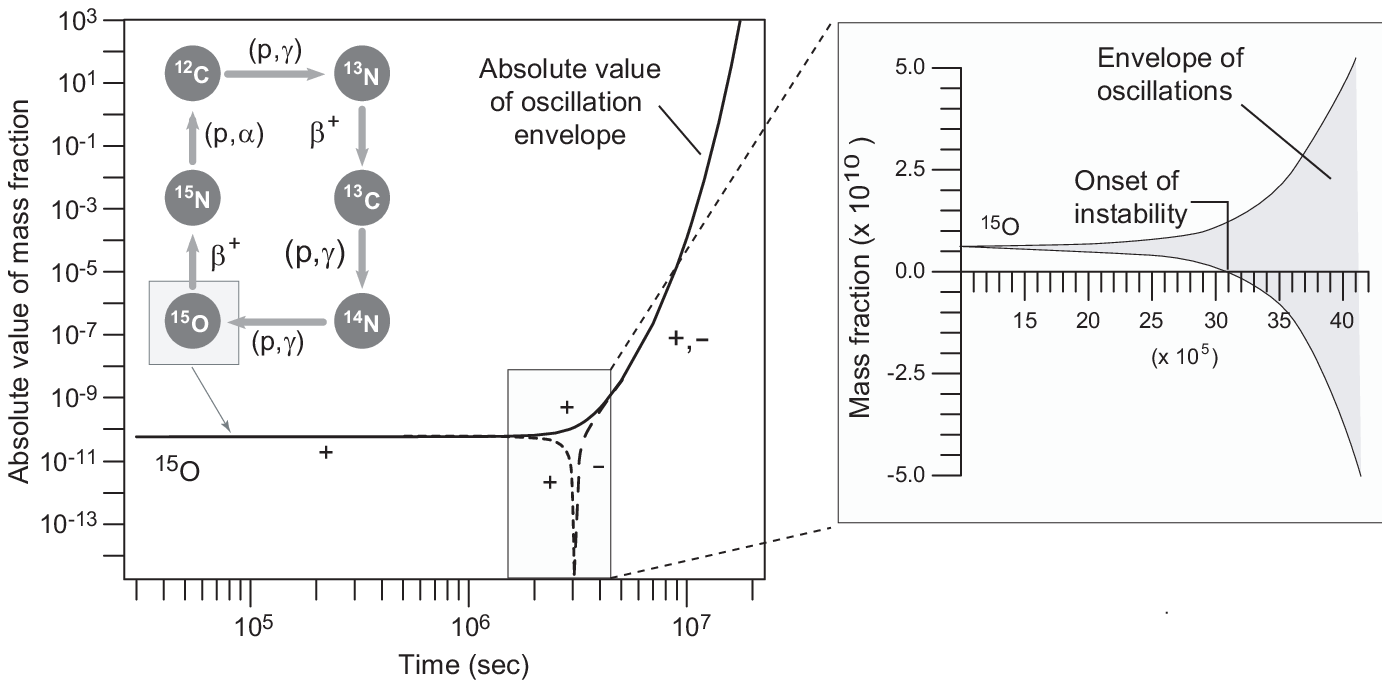}
{Origin of the stiffness instability for explicit Euler-method integration of
the CNO cycle main branch (shown as an inset).  If the timestep is too large the
mass fraction of $^{15}$O, which normally is of order $10^{-10}$ or smaller
under the conditions assumed here, can become negative.  This unphysical
condition is unstable and triggers an exponential runaway that quickly crashes
the entire network.  On the left side, the {\em absolute value} of the envelope
of the oscillating solution is plotted on a log scale, with the sign indicated
adjacent to different parts of the curves. The right side shows a blowup of the
region where the mass fraction begins to oscillate into negative values, now
plotted on a linear scale to show clearly the envelope of the diverging
oscillating solution. This figure is a generalization of \fig{negProb} to a more
complex network. }
applies the preceding model of the origin of the stiffness instability to the
CNO cycle.  As noted earlier, (for the parameters assumed) ordinary explicit
Euler integration is highly unstable for a timestep larger than 285.7 seconds.
This critical timestep is exactly two divided by the fastest rate parameter in
the system, which is that for the $\beta$-decay of $^{15}$O to $^{13}$N.
Expressing the network in matrix form and diagonalizing indicates that the
origin of this instability lies in eigenvalues that exceed unity in absolute
value if $h>2/k\tsub{max}$.  Generalizing the discussion associated with Eq.\
(\ref{negative1.2}), for a coupled set of equations a finite-difference
iteration for $m$ steps entails a matrix raised to the power $m$ applied to the
original $y$ vector.  This is guaranteed to converge only if no eigenvalue of
the matrix exceeds unity in magnitude.

Less abstractly, Fig.\  \ref{fig:O15LogAbsoluteDiverge} indicates that the
origin of this instability for standard explicit integration is the tendency of
the $^{15}$O population to become negative for large explicit timesteps. This is
a more complex system than the previous simple decaying-exponential example
because the $^{15}$O population is depleted by the $\beta$-decay but also
replenished by the $(p,\gamma)$ proton capture reaction on $^{14}$N (see Fig.\
\ref{fig:cnoCycle} and the inset diagram on the left side of Fig.\
\ref{fig:O15LogAbsoluteDiverge}).  Nevertheless, we see that the origin of the
stiffness instability is very similar to that illustrated in the preceding
simple example: a small box population becomes unphysically negative because of
taking too large a timestep, so in the next timestep the offending box exports
an unphysical flux of negative population, triggering a divergence that rapidly
compromises the entire network. 

Motivated by the properties of stiffness illustrated in the previous examples,
we first introduce a rather crude approximation for stabilizing explicit
integration in stiff networks.  Although we shall discuss  much better
algorithms after that, it is instructive that even this simple approximation, by
removing a primary source of stiffness, leads to an explicit algorithm that is
usable in large, realistic, extremely-stiff networks.

\section{\label{sh:flfdAlgorithm} Flux-Limited Forward Difference Algorithm}

The results of the previous section indicate that one form of stiffness
instability is generated by box populations that become very slightly negative
and that these destabilize the network when they are exported as fluxes to other
boxes (effectively because this anomalous sign for the box population turns what
should be exponential decay into exponential growth).  Thus, we invoke a simple
flux-limiting prescription that if the population of a box becomes negative we
do not change the population itself (which would quickly violate conservation of
probability) but we suppress all export of that negative population to other
boxes.  Thus, in the course of a calculation no flux is permitted out of boxes
that have negative populations until their populations again become positive
because of flux into them.  Formally, we require for all computed outgoing
fluxes that $F_{ij} \rightarrow \max (F_{ij},0) \equiv \tilde F_{ij}$. In
shorthand, we refer to this as {\em suppression of negative flux,} and refer to
the resulting algorithm as the {\em flux-limited forward difference (FLFD)
algorithm.} In its simplest implementation, which we shall illustrate here, we
use the (explicit) forward Euler method supplemented by the flux-limiting
prescription, but the same approach can be applied to higher-order forward
differencing.

Because we do not alter populations but only restrict their flow by this
algorithm, it conserves probability.  We may expect that some populations in the
network will now be in error because of the flux-suppression criterion. However,
such effects tend to involve the smallest populations (because they are the ones
most easily made negative by the numerical error), so we may expect that this
approximation could be a good one for the larger populations, with the error
concentrated in the smallest populations.  For large networks coupled to
hydrodynamical evolution it is the changes in the larger populations that
dominate energy production and concentration changes, so errors in the smaller
populations are largely irrelevant. Therefore, an algorithm that removes
stiffness by preferentially concentrating errors in the smallest populations may
be a very usable one. In the next section we test this idea on a realistic
thermonuclear network.

\section{\label{sh:novaExample} Flux-Limited Network Solutions under Nova
Conditions}

We illustrate  the flux-limited forward difference algorithm by application to
astrophysical thermonuclear networks under nova conditions.

\subsection{Hot CNO Burning}

Some representative isotopic abundances under nova (hot CNO cycle) conditions
calculated using the FLFD algorithm are shown in Fig.~\ref{fig:compare}.%
\putfig
{compare}
{0pt}
{\figdn}
{0.85}
{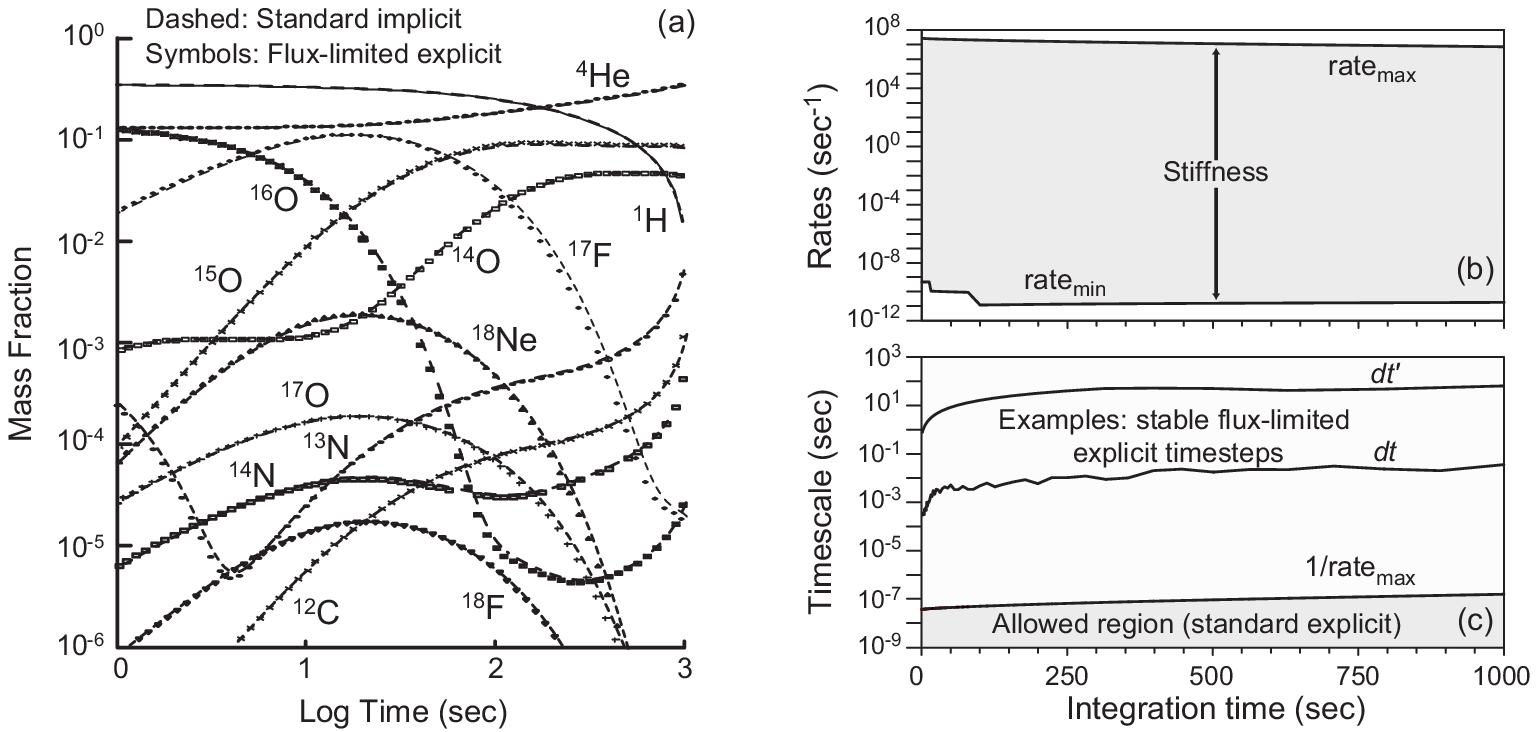} 
{(a)~Some representative isotopic abundances under nova (hot CNO) conditions
calculated using the explicit FLFD algorithm and compared with a calculation
using the standard implicit solver Xnet \cite{raphcode}. A constant temperature
of $T=0.25 \times 10^9$ K and constant density $\rho = 500 \units{g\,cm}^{-3}$
were assumed. The explicit network contained 145 isotopes, with 924 non-zero
couplings. (b)~Rates and timescales characteristic of a FLFD nova simulation.
Conditions as for part (a) but with a larger reaction library: 896 isotopes with
8260 couplings were included (though only several hundred isotopes were
populated significantly). Extremal rates plotted are restricted to those
involving non-zero fluxes. (c)~Comparison of maximum stable timestep [$\simeq
1/{\rm rate}_{\rm\scriptstyle max}$ from part (b)] possible for a standard
explicit integration with much larger stable timesteps $dt$ and $dt'$ for some
representative explicit FLFD integrations.}
An initial isotopic abundance distribution enriched in heavy elements has been
assumed \cite{parete-koon03}, and reaction rates from the REACLIB library
\cite{raus2000} have been used. Figure~\ref{fig:compare}(a) displays some
representative populations, with the results of a standard implicit
calculation~\cite{raphcode} shown as dashed lines and FLFD calculations shown as
symbols.  An adaptive timestep was used in the FLFD integration, with the
timestep adjusted to keep the populations transferred between boxes in a
timestep for some key populations within a prescribed range. This timestep is
{\em much} larger than would be stable in a standard explicit integration, as we
shall discuss further in \S\ref{ss:stiffness}. Note the very good agreement
between implicit and explicit flux-limited methods over six orders of magnitude
in the mass fractions in Fig.~\ref{fig:compare}(a). In a realistic coupling of
such a network to hydrodynamics in a nova simulation, isotopes with mass
fractions smaller than $10^{-2}$--$10^{-3}$ would likely be not very important
to the hydrodynamical evolution.

\subsection{\label{ss:stiffness} Stiffness and Stability}

Figure~\ref{fig:compare}(b) displays the  fastest and slowest rates entering a
representative FLFD nova simulation as a function of time. The difference of
some 18 orders of magnitude between the fastest and slowest rates at any
timestep is an indication that this is an extremely stiff system. For standard
explicit algorithms, the largest timestep permitted by stiffness stability
criteria generally is of order the inverse of the fastest rate in the network
(see the discussion in \S\ref{ss:simpleMod} and in Ch.~16 of Ref.\
\cite{press92}).  For the calculations illustrated in Fig.~\ref{fig:compare}(b),
the inverse of the fastest rate gives the lower curve in
Fig.~\ref{fig:compare}(c).  Thus a normal explicit algorithm would be restricted
by stability requirements to timesteps lying approximately in the shaded region
below this curve ($dt \simeq 10^{-7}$ seconds or less).

In contrast, Fig.~\ref{fig:compare}(c) displays two curves for stable FLFD
integration  timesteps lying far above this region.  The curve marked $dt$ is
for a timestep small enough to give accuracy comparable to
Fig.~\ref{fig:compare}(a).  This timestep is  seen to be about $10^5$ times
larger than would be stable for a normal explicit integration.  The curve marked
$dt'$ is for a much larger FLFD algorithm timestep that compromises accuracy for
the weaker transitions but remains stable and calculates the stronger
transitions correctly.  The timestep $dt'$ $\sim$ 100 seconds is about $10^9$
times larger than would be stable for a standard explicit algorithm.  Since
$dt'$ already is comparable to the characteristic timescale of a nova explosion,
this example exhibits a stable explicit integration timestep for a realistic,
extremely stiff system that is effectively arbitrarily large with respect to the
usual upper limit for explicit integration.

The picture that emerges from the present results  is that the FLFD method is
susceptible to stiffness instability, just as for any other explicit method. 
However, the effect of the instability can be confined to controlled errors in
small populations for a range of timesteps far beyond the normal onset of
stiffness instability simply by imposing a flux limiter.  This flux limiter does
not prohibit negative populations but prevents their export between boxes in the
network and thus prevents them from growing in uncontrolled fashion.

\section{\label{sh:equil-nonequil} Stiffness under Equilibrium and
Non-Equilibrium Conditions}

The results of the preceding section indicate that even a relatively crude limit
on propagation of negative population can remove large amounts of stiffness and
leads to an algorithm useful in realistic applications, even for the extremely
stiff networks common in astrophysics.  But we now will demonstrate that one can
do much better than that, by exploiting the algebraic structure of the
differential equations to make a better approximation than just constraining the
sign of propagating populations. 

The first step is that we must look more deeply at the stiffness instability. As
we now discuss, for the sort of large and very stiff networks that we are
addressing here there are several fundamentally different sources of stiffness
instability that are often not clearly distinguished in the literature.  The
examples discussed to this point (and in many textbooks) emphasize the type of
instability associated with small quantities that should strictly be
non-negative becoming negative because of an overly ambitious numerical
integration step.  The discussion of the flux-limited forward difference
algorithm in \S\ref{sh:flfdAlgorithm} illustrates that this type of instability
can be removed by approximations that do not permit unphysical negative
quantities to influence the rest of the network. However, there are other
stiffness instabilities that may be initiated even when no population variables
become negative in an integration step.  In this kind of instability we end up
having to take the difference of large numbers to obtain a result very near
zero.  The numerical errors that ensue in a standard explicit approach can then
accumulate rapidly and destabilize the network, even before any abundances
become negative. This may still be viewed as a stiffness instability because it
results from a numerical integration trying to deal with very different
timescales, but the origin of these timescales is different from that discussed
above. In this case the disparate timescales are the very rapid reactions
driving the system to equilibrium contrasted with the very slow timescale
associated with equilibrium itself (which tends to infinity).

As we now consider, this distinction  is essential to what follows because these
stiffness instabilities have essentially different solutions. Furthermore, we
shall find that the second kind of instability can be divided into two
subclasses requiring different stabilizing approximations, and that the
approximations that we shall introduce in these cases will also take care
naturally of the first class of stiffness instabilities because they will as a
matter of course prevent the occurrence of negative probabilities in the
network.

\subsection{\label{ss:approachEquil} The Approach to Equilibrium}

We shall use ``equilibrium'' in a broad sense to mean a condition where  the
populations in a network are being strongly influenced by competition between
terms of opposite sign on the right sides of the differential equations
governing their evolution. How do we measure the degree of equilibration in a
large, stiff network?  In terms of the coupled set of differential equations
describing the network, we may distinguish two qualitatively different
conditions:

\begin{enumerate}
 \item 
A macroscopic equilibration that acts at the level of individual differential
equations.
\item
A microscopic equilibration that acts at the level of individual terms within a
given differential equation.
\end{enumerate}
Let us consider each of these cases in turn. The differential equations that we
must solve take the general form of \eq{eq1.1}, $dy_i/dt = \fplus i - \fminus
i$, where the total flux has been decomposed into  a component $\fplus i$
increasing the population $y_i$ and a component $\fminus i$ depleting the
population $y_i$ in a given timestep.

\subsubsection{\label{ss:macroscopic} Macroscopic Equilibration}

One class of approximations that we will investigate depends upon assuming that
$\fplus i - \fminus i \rightarrow 0$ (asymptotic approximations) or $\fplus i -
\fminus i \rightarrow$  constant (steady-state approximations).  We shall refer
to these conditions as a macroscopic equilibration, since they involve the
entire right side of a differential equation in \eq{eq1.1} tending to zero or a
finite constant. We shall introduce approximations exploiting this whereby
whole differential equations are removed from the numerical integration for a
network timestep in favor of a algebraic approximate solutions for that
timestep. Such approximations don't reduce the number of equations to integrate,
but they reduce the number of equations integrated {\em numerically
by forward difference}.  They reduce stiffness for any remaining equations that
are integrated numerically in the timestep because removing the equations
satisfying these conditions tends to reduce the disparity in timescales for the
remaining equations.

\subsubsection{\label{ss:microscopic} Microscopic Equilibration}

In \eq{eq1.1}, $\fplus i$ and $\fminus i$ for a given species $i$ generally each
consist of a number of terms depending on the other populations in the network, 
\begin{eqnarray}
\deriv{y_i}{t} &=&
\fplus{i} - \fminus{i} 
\nonumber
\\
&=& (f_1^+ + f_2^+ + \ldots)_i -  (f_1^- + f_2^- + \ldots)_i
\nonumber
\\
&=& (f_1^+
- f_1^-)_i + (f_2^+ - f_2^-)_i + \ldots = \sum_j (f^+_j - f^-_j)_i,
\label{equilDecomposition}
\end{eqnarray}
At the more microscopic level, groups of individual terms on the right side of
\eq{equilDecomposition} may come approximately into equilibrium (the sum of
their fluxes tends to zero), even if macroscopic equilibration conditions are
not satisfied. The simplest possibility for this {\em microscopic
equilibration} is that forward--reverse reaction pairs such as $A+B
\rightleftharpoons C$, which will contribute flux terms with opposing signs on
the right sides of differential equations in which they participate, come
approximately into equilibrium. Then we may consider an algebraic approximation
that removes groups of such terms from the numerical integration, replacing
their sum of fluxes identically with zero. This will not generally reduce the
number of equations to be integrated numerically in a network timestep, but it
can reduce dramatically the stiffness of those equations by removing terms with
fast rates from the equations, thereby reducing the disparity between the
fastest and slowest timescales in the system.

Such considerations will be the basis of the partial equilibrium methods that
will be discussed in depth in the third paper in this series \cite{guidPE}.
There we shall also demonstrate two important general conclusions:
(1)~Approximations based on microscopic equilibration are much more efficient at
removing stiffness than those based on macroscopic equilibration, because they
target more precisely the sources of stiffness in the network.  (2)~The most
powerful approach will be to use macroscopic and microscopic approximations
simultaneously in the same set of equations, because they can complement each
other in removing stiffness from the equations to be integrated numerically.

\subsection{\label{ss:measureEquil} A Quantitative Measure of
Microscopic Equilibration}

The preceding section suggests that when microscopic equilibration becomes
important in a network, the methods for dealing explicitly with stiffness are
different from those used to deal with macroscopic equilibration.  Therefore, it
is important to establish a quantitative measure of how much microscopic
equilibration is present in the network.  We shall introduce the simplest
possibility: that the amount of microscopic equilibration in the network is
measured by the fraction of reaction pairs such as $A+B+\ldots
\rightleftharpoons C+D+\ldots$ that are judged to be in equilibrium (with each
reaction pair considered in isolation from the rest of the network for purposes
of this determination).

The full machinery to carry this out will be described in the third paper of
this series \cite{guidPE}, but our essential approach will be to extend the
formalism introduced by Mott \cite{mott99} to derive conditions on the
populations for equilibrium in a reaction pair, and to determine whether the
pair is in equilibrium by comparing the actual populations computed from the
preceding network timestep with the theoretical equilibrium populations.  The
general result that we shall obtain is that for a forward--reverse reaction
pair of the form $A+B+\ldots \rightleftharpoons C+D+\ldots$, the
equilibrium abundance $\bar y_i$ of each species is given by
\begin{equation}
 \bar y_i \equiv y^{{\rm\scriptstyle eq}}_i = -\frac{1}{2a} 
(b + \sqrt{-q}).
\label{2body1.6}
\end{equation}
where $q \equiv 4ac-b^2$, the parameters $a$, $b$, and $c$ are known
functions of the current rate
parameters and the populations at the beginning of the timestep, 
and the approach to equilibrium for the reaction pair
is governed by a single timescale $\tau = q^{-1/2}$.  We may then estimate
whether a given reaction is near equilibrium at time $t$ by requiring
\begin{equation}
 \frac{| y_i(t) - \bar y_i |}{\bar y_i}
< \epsilon_i
\label{2body1.8b}
\end{equation}
for each species $i$ involved in the reaction, where $y_i(t)$ is the actual
abundance, $\bar y_i$ is the equilibrium abundance determined by \eq{2body1.6},
and $\epsilon_i$ is a user-specified tolerance that we shall choose typically
to be of order $10^{-2}$. Alternatively, we may use the equilibration timescale
$\tau$ compared with the numerical timestep being taken as a measure of
microscopic equilibration.

We shall term a network {\em strongly (microscopically) equilibrated} if a
significant fraction of its reaction pairs satisfy the condition
(\ref{2body1.8b}) or its equilibration timescale $\tau$ is much less than than
the current integration timestep, and {\em weakly (microscopically)
equilibrated} if at most a few percent of its reaction pairs satisfy Eq.\
(\ref{2body1.8b}) or have values of $\tau$ considerably larger than the
integration timestep. The remainder of this paper will deal with methods based
on asymptotic approximations to stabilize explicit integration for networks that
are at most weakly equilibrated, and a second paper will deal with methods based
on quasi-steady-state approximations to stabilize weakly-equilibrated systems
\cite{guidQSS}. The corresponding stabilization of networks near microscopic
equilibrium will be discussed in the third paper of this series \cite{guidPE}.

\section{\label{algebraic} Algebraic Stabilization of Solutions Using
Asymptotic Approximations}

  The flux-limited forward difference approximation described in
\S\ref{sh:novaExample} illustrates the basic principle that explicit integration
in non-equilibrium situations can be stabilized by forbidding the propagation of
negative flux.  However, the FLFD algorithm represents only a zero-order
solution to this problem (a yes/no decision on whether flux components are
allowed to propagate) that can be improved substantially by exploiting the
structure of the coupled equations to replace the numerical solution with an
algebraic approximation that is strictly non-negative.  Although the
approximations that we now discuss have been implemented in some form in earlier
literature \cite{oran05,youn77,mott00,mott99}, we shall find that our
implementation appears to be much more successful than previous applications to
large, extremely stiff networks, and we shall reach different conclusions about
these methods than those reached in earlier publications.

\subsection{\label{ss:asymptoticSimple} Some Explicit Asymptotic Approximations}

The differential equations that we must solve take the  form given by
\eq{eq1.1}. Generally, $\fplus i$ and $\fminus i$ for a given species $i$ each
consist of a number of terms depending on the other populations in the network.
The depletion flux for the population of species $i$
will be proportional to $y_i$,
\begin{equation}
    \fminus i = (k_1^i + k_2^i + \ldots + k_m^i)y_i \equiv k^iy_i,
\label{eq1.2}
\end{equation}
where the $k_j^i$ are rate parameters (in units of time$^{-1}$) for each of the
$m$ processes that can deplete $y_i$, which may  depend on the populations $y_j$
and on hydrodynamical variables such as temperature and density. The
characteristic
timescales $\tau^i_j = 1/k^i_j$ will differ by many orders of
magnitude in the systems of interest, implying that the equations are 
stiff. From \eq{eq1.2} we may define the effective total depletion rate $k^i$
for  $y_i$ at a given time, and a corresponding timescale $\tau^i$ as
\begin{equation}
    k^i \equiv  \frac{\fminus i}{y_i}
\qquad
\tau^i = \frac{1}{k^i},
\label{eq1.3}
\end{equation}
permitting \eq{eq1.1} to be written as
\begin{equation}
    y_i = \frac{1}{k^i} \left( \fplus i - \deriv{y_i}{t} \right).
\label{eq1.4}
\end{equation}
Thus, in a finite-difference approximation at timestep $t_n$ we have
\begin{equation}
    y_i(t_n) = \frac{\fplus i (t_n)}{k^i(t_n)} - \frac{1}{k^i(t_n)} \left.
\deriv{y_i}{t}
    \right|_{t=t_n}.
\label{eq1.5}
\end{equation}
We now define the {\em asymptotic limit} for the species $i$ to be $\fplus i
\simeq \fminus i$, implying from \eq{eq1.1} that $\diffelement y_i /\diffelement
t \simeq 0$.  In this limit \eq{eq1.5} gives a first approximation 
$ y^{(1)}_i (t_n)$ and local
error $E_n^{(1)}$, respectively, for
$y_i(t_n)$ as
\begin{equation}
 y^{(1)}_i (t_n) = \frac{\fplus i(t_n)}{ k^i(t_n)}
\qquad
E_n^{(1)} \equiv y(t_n) - y^{(1)}(t_n) = -\frac{1}{k(t_n)}
\deriv yt(t_n)
\label{firstApprox}
\end{equation}
For small $\diffelement y_i/\diffelement t$ we may then get a correction 
term by writing the derivative term in \eq{eq1.5} as
\begin{equation}
 \deriv{y}{t}(t_n) = \frac{1}{\Delta t}
    \left( y_i(t_n) - y_{i}(t_{n-1}) \right)
+\frac{1}{\Delta t} \left( E_n^{(1)} - E_{n-1}^{(1)}\right) +  O(\Delta t) ,
\label{derivative}
\end{equation}
where $O(x)$ denotes order of $x$.  If we retain only the first term and
approximate $y(t)$ by $y^{(1)}(t)$, we obtain
\begin{eqnarray}
    \left. \deriv{y}{t}\right|_{t=t_n} &\simeq& \frac{1}{\Delta t}
    \left( y_i^{(1)}(t_n) - y_{i}^{(1)}(t_{n-1}) \right)
= \frac{1}{\Delta t} \left( \frac{\fplus i(t_n)}{k^i(t_n)} 
    -\frac{\fplus i (t_{n-1})}{k^i(t_{n-1})} \right).
\label{eq1.7}
\end{eqnarray}
Therefore, 
the estimate for $y$
is improved to
\begin{equation}
    y_n^{(2)} \simeq \frac{F^+_n}{k_n} - \frac{1}{k_n\Delta t}
     \left(\frac{F^+_n}{k_n} - \frac{F^+_{n-1}}{k_{n-1}} \right),
\label{asySimple}
\end{equation}
where we now employ compact index notation $y_n \equiv y_i(t_n)$, and so on, and
have dropped the species index $i$ to avoid notational clutter. Because we are
approximating the derivative term, we expect that \eq{asySimple} is valid only
if the second term is small, implying that our approximation becomes more valid
if $k\Delta t$ is large. 

The preceding discussion implements an asymptotic approximation in a very simple
way.  Other more sophisticated asymptotic approximations may be derived. For
example, if we retain the full expression for the derivative in \eq{derivative}
and substitute in \eq{eq1.5}, we obtain
$$
y^{(2)}_n = \frac{F^+_n}{k_n} -\frac{1}{k(t_n) \Delta t} \left( y(t_n) -
y(t_{n-1}
\right) 
- \frac{1}{k(t_n) \Delta t} \left( -E_n^{(1)} + E_{n-1}^{(1)}\right).
$$
Setting $y(t_n) = y^{(2)}_n$ and solving for $y^{(2)}_n$ gives
\begin{equation}
y_n^{(2)} = \frac{1}{1+k_n\Delta t} \left(y_{n-1} + F^+_n \Delta t \right),
\label{asySophia}
\end{equation}
where a term of order $(\Delta t) ^2/(1+k(t_n)\Delta t)$ has been discarded.
Another approach is to use a predictor--corrector scheme
\cite{oran05,mott00,mott99}. We may solve
\eq{eq1.1} by finite difference, replacing quantities on the right side of
\eq{eq1.1} with averaged quantities:
\begin{equation}
 \frac{y_n - y_{n-1}}{\Delta t} = \tfrac12 (F^+_n + F^+_{n-1})
-\frac{y_n+ y_{n-1}}{\tau_{n} + \tau_{n-1}},
\label{finiteAvg}
\end{equation}
where for notational convenience we have traded the rate parameters $k_i$
for the corresponding timescales $\tau_i \equiv 1/k_i$. Solving this equation
for $y_n$ gives
\begin{equation}
 y_{n} = \frac{(\tau_n + \tau_{n-1} - \Delta t)y_{n-1} + \tfrac12 
(\tau_n + \tau_{n-1})(F^+_n + F^+_{n-1})\Delta t}
{\tau_n + \tau_{n-1} + \Delta t} ,
\label{soln1}
\end{equation}
which is not easy to use because the $\tau_n$ and $F^+_n$ are implicit functions
of the $y_n$ that we seek. We obtain a usable explicit algorithm by solving an
approximate form of \eq{soln1} to get a first guess $y\tsup p$ for $y_n$ (the
predictor step), using that approximate result to estimate values of $\tau_n$
and $F^+_n$, and then using these values obtained from the predictor step to
solve \eq{soln1} (the corrector step).  Specifically,
\begin{itemize}
 \item 
The predictor estimate $y\tsup p$ is obtained by setting $\tau_n = \tau_{n-1}
\equiv \tau_0$
and $F^+_n = F^+_{n-1} \equiv F^+_0$ in \eq{soln1}.
\item
The corrector value
$y\tsup c$ is
obtained by substituting  $F_{\scriptstyle\rm p}^+$ for
$F^+_n$, and
$\tau\tsub p$ for $\tau_n$ in \eq{soln1}.
\end{itemize}
This yields a predictor and corrector \cite{oran05}
\begin{eqnarray}
 y\tsup p &=& 
\frac{y^0(2\tau_0-\Delta t)+2 F_0^+\tau_0 \Delta t}
{2\tau_0 + \Delta t} \qquad \units{(Predictor)}
\nonumber
\\
y\tsup c &=& \frac{y^0(\tau\tsub p + \tau_0 - \Delta t) + \tfrac12
\Delta t
(F_{\scriptstyle\rm p}^+ + F_0^+)(\tau\tsub p + \tau_0)}
{\tau\tsub p + \tau_0 + \Delta t} \qquad \units{(Corrector)}
\label{asyOB}
\end{eqnarray}
where ``0'' denotes initial quantities and ``p'' denotes quantities computed
using the results of the predictor step.  We shall test the asymptotic
approximations \eqnoeq{asySimple}, \eqnoeq{asySophia}, and \eqnoeq{asyOB} in
examples below.

\subsection{\label{ss:asymptoticProperties} Mathematical Properties of
Asymptotic Approximations}

The mathematical and numerical properties of asymptotic approximations have been
explored in Refs.\ \cite{oran05,youn77,mott00,mott99} and we only summarize
them here. 

\begin{enumerate}
 \item 
The preceding asymptotic formulas are explicit,
since only quantities already known are required to advance a timestep.
\item
Asymptotic methods do not need to compute Jacobians or
invert matrices.
\item
Asymptotic methods are A-stable \cite{oran05,dahl63} on linear problems.
\item
Explicit asymptotic
methods should scale  linearly with network size.
\item 
Asymptotic methods require initial values from only one timestep
(self-starting).
\end{enumerate}
A more extensive discussion may be found in Oran and Boris \cite{oran05} and in
Mott \cite{mott99}.

\subsection{\label{ss:asyAlg} An Asymptotic Flux-Limiting Algorithm}

We now use the preceding formalism to define an explicit asymptotic
flux-limiting integration algorithm.  Since the asymptotic approximation
specified above is expected to be valid if $k\Delta t$ is large, we define a
critical value $\kappa$ of $k\Delta t$ and at each timestep cycle through all
network populations and compute the product $k^i\Delta t$ for each species $i$
using \eq{eq1.3} and the proposed timestep $\Delta t$.   Then, for each
population species $i$
\begin{enumerate}
\item
If $k^i\Delta t < \kappa$, we update the population numerically using the
standard flux-limiting explicit algorithm discussed in \S\ref{sh:flfdAlgorithm}.
\item
Otherwise, for $k\Delta t \ge \kappa$, we update the population algebraically
using one of the asymptotic approximations given in Eqs.\
\eqnoeq{asySimple}, \eqnoeq{asySophia}, or \eqnoeq{asyOB}.
\end{enumerate}
From considerations such as those in \S\ref{sh:large-stiff} we expect an
explicit integration to be stable if $k^i\Delta t
<1$ and to potentially be unstable if $k^i\Delta t \ge 1$, a possible choice is
$\kappa = 1$.  For the networks discussed in this paper, we have found this
value of $\kappa$ to work well and have adopted it. For that specific choice
of $\kappa$, explicit numerical integration should be stable for those cases
where it is applied, so we expect that at each timestep all abundances will
remain positive and the flux-limiting prescription for the explicit numerical
integration may be dropped. Notice that at each timestep some species may be
updated by explicit forward difference and some by the asymptotic approximation,
and that the division of species between these two categories could change at
each timestep since the product $k^i\Delta t$ is time-dependent.

\subsection{\label{ss:asyTimestep} A Simple Adaptive Timestepper}

Implementing the preceding algorithm requires an adaptive timestepper.  We take
the point of view that the present task is to establish whether explicit methods
can even compete with implicit methods for stiff networks.  Since previous
studies have generally found that explicit methods fail by many orders of
magnitude to attain the speeds of implicit methods in highly-stiff networks, our
timestepper need not be highly optimized at this point to answer that question. 

The first consideration is a standard one that limits the population change at
each timestep. However, the preceding algorithm does not guarantee that total
population is conserved, so in setting an adaptive timestep one should make a
check that ensures conservation of population at the desired level. Generally,
if population conservation does not satisfy the required tolerance at a given
timestep, making the timestep small enough is guaranteed to improve conservation
of population because it will reduce $k^i\Delta t$ and therefore will tend to
decrease the number of isotopes being treated asymptotically. To ensure
conservation of particle number at a desired level in the overall calculation,
we may limit the deviation in any one timestep to a small amount. Thus, we adopt
a simple timestepper with two stages:
\begin{enumerate}
 \item 
After the rates and fluxes are computed at the beginning of a new timestep,
compute a trial timestep based on limiting the change in populations that would
result from that timestep to some specified tolerance. Use the minimum of this
trial timestep and the timestep that was taken in the previous integration
timestep to update the populations by the explicit asymptotic algorithm.
\item
Check for conservation of population. If the conservation law is satisfied
within the desired tolerance for this timestep, proceed.  If it is not, or is
satisfied too well, decrease or increase the timestep as appropriate by a small
factor and repeat the calculation of populations with the new timestep but
original fluxes. 
\end{enumerate}
We then accept this timestep, without further check. One could also iterate  to
ensure that the conservation condition is satisfied at each step, but this did
not significantly improve the results in our tests.

Far-removed from equilibrium the limitation in population changes 
determines the timestep with this algorithm, but in the approach to equilibrium
the timestep becomes dominated by the probability conservation
criterion. Though it is likely not very optimized, we have found this simple
timestepper to be stable and accurate for the varied astrophysical thermonuclear
networks that we have tested, and thus adequate for our task here. This should
be contrasted with previous attempts to apply asymptotic methods to
thermonuclear networks, which failed to produce accurate results and were
abandoned as unsuitable for such stiff networks \cite{oran05,mott99}. 

% See Oran and Boris, p. 147 for failure of asymptotic and QSS for
% thermonuclear above

\section{Comparisons of Explicit and Implicit Integration Speeds}

We shall be comparing explicit and implicit methods using codes that are at very
different stages of development and optimization. We assume that for codes at
similar levels of optimization the primary difference between explicit and
implicit methods  would be in the extra time spent in implicit-method matrix
operations. Hence, if the fraction of time spent on linear algebra is $f$ for an
implicit code, we assume that an explicit code at a similar level of
optimization could compute a timestep a factor of $F = 1/(1-f)$ faster. In Table
\ref{tb:explicitSpeedup} we display factors $F$
\begin{table}
\caption{\label{tb:explicitSpeedup}Explicit speedup factors $F$}
\begin{indented}
\item[]\begin{tabular}{@{}lll}
\br
Network & Isotopes & Speedup $F$\\
\mr
            pp &
            6 &
            $\sim 1.5$

        \\ 
            Alpha &
            16 &
            3

        \\
            Nova &
            134 &
            7

        \\
            150-isotope &
            150 &
            7.5

        \\
            365-isotope &
            365 &
            $\sim 20$

        \\        
\br       
\end{tabular}
\end{indented}
\end{table}
based on data obtained by Feger \cite{feg11a,feg11b} using the implicit,
backward-Euler code Xnet \cite{raphcode} with both dense and sparse solvers.
Then we may compare roughly the speed of explicit versus implicit codes
(possibly at different levels of optimization) by multiplying $F$ by the ratio
of implicit to explicit integration steps required for a given problem.  This
procedure has obvious uncertainties, and likely underestimates the speed of an
optimized explicit versus optimized implicit code \cite{guidPE}, but will give a
useful lower limit on how fast the explicit calculation can be.

\section{\label{sh:asyExamples} Network Calculations in Simplest Asymptotic
Approximation}

First we shall establish that the asymptotic algorithm is capable of correct
integration of coupled sets of extremely stiff equations.  We have tested this
for a variety of calculations in two ways:  

\begin{enumerate}
 \item 
Comparisons with results from standard implicit codes, and 
\item
For some smaller networks where the corresponding integration time is not
prohibitive, comparison with explicit forward Euler calculations made with
timesteps short enough to be stable.
\end{enumerate}
Our general finding is that the asymptotic algorithm outlined above gives the
same results as standard implicit and explicit codes, even for the stiffest
networks found in astrophysics applications, provided that the numerical
timesteps are limited sufficiently to ensure conservation of overall probability
in the network at the desired level of precision.

\subsection{\label{explicitTest} Tests Against Fully-Explicit Calculations}

Figure \ref{fig:compareAsyExplicitT9_7rho1e8}%
\singlefig
{compareAsyExplicitT9_7rho1e8}
{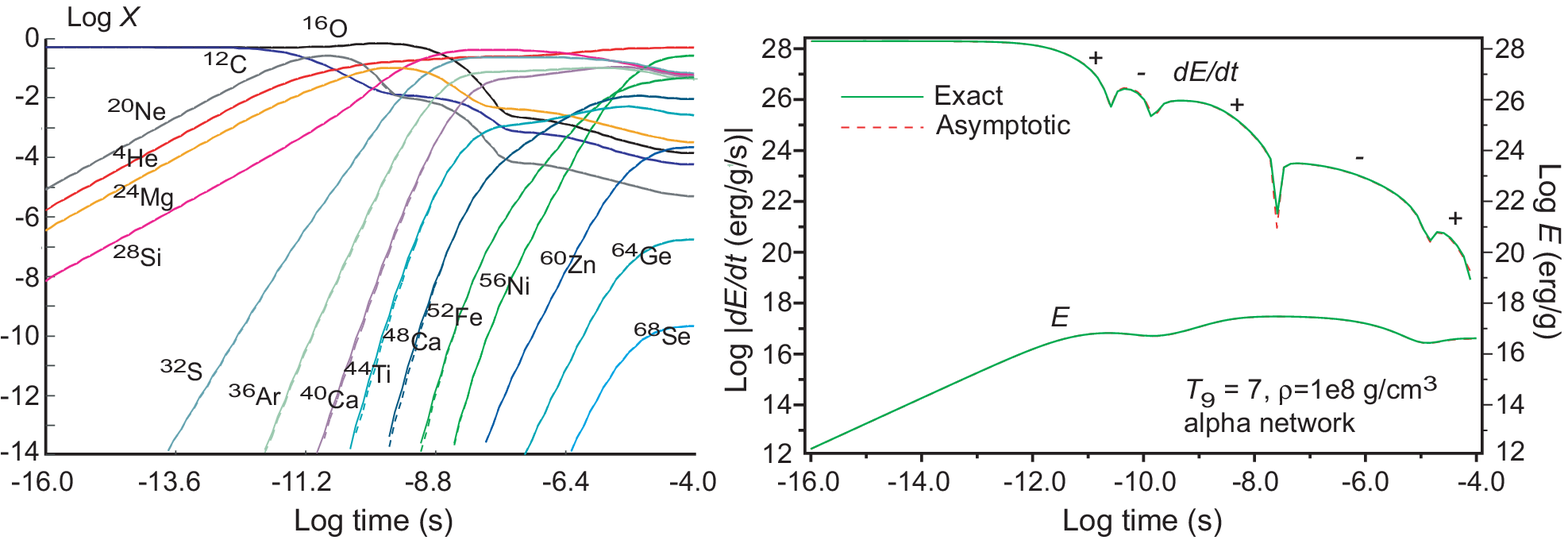}
{0pt}
{\figdn}
{0.69}
{Comparison of the asymptotic approximation (\ref{asySimple}) with a forward
Euler integration that used timesteps short enough to be stable for an alpha
network at constant temperature $T_9=7$ and density of $10^8
\units{g\,cm}^{-3}$. Initial abundances corresponded to equal mass fractions
of \isotope{12}C and \isotope{16}O, and rates from the REACLIB library
\cite{raus2000} were used. The left figure compares mass fractions; the right
figure compares differential and integrated energy production. Solid lines are
explicit forward-Euler integration with a timestep constrained to be less than
the inverse of the fastest rate in the system so that it is stable; dashed lines
are the corresponding asymptotic approximation using \eq{asySimple}. The value
of $dE/dt$ oscillates between positive and negative values so the log of the
absolute value of $dE/dt$ is plotted, with the sign of $dE/dt$ indicated in each
region. }
illustrates an asymptotic approximation calculation that gives results
essentially identical to results from exact numerical integration. In this
example the sum of the mass fractions was constrained to deviate from unity by
not more than 1\% over the entire range of the calculation by requiring that if
it deviates by more than $10^{-8}$  from unity in any one timestep, the timestep
is reduced in size. We shall generally use this global 1\% criterion for
examples presented in this paper. Higher precision may be obtained by tightening
this constraint, but this is typically already more conservative than justified 
since input parameter uncertainties in realistic large thermonuclear networks
and uncertainties in the coupled hydrodynamics may each be considerably larger
than 1\%. 

Having established that explicit asymptotic approximations can give correct
results even for extremely stiff networks, we now turn to the question of their
efficiency (and further tests of accuracy) by examining calculations of various
astrophysical thermonuclear networks using this approximation. In this section
we shall use the simplest asymptotic approximation, corresponding to Eq.\
(\ref{asySimple}). In \S\ref{sh:asySophisticated} we shall test the alternative
asymptotic approximations of Eqs.\ (\ref{asySophia}) and (\ref{asyOB}).

\subsection{Explicit Asymptotic Integration of the pp-Chains}
\protect\label{ppAsy}

The solar pp-chains  provide a striking example of stiffness in a simple
network. In \fig{ppChains}%
 \singlefig
     {ppChains}
     {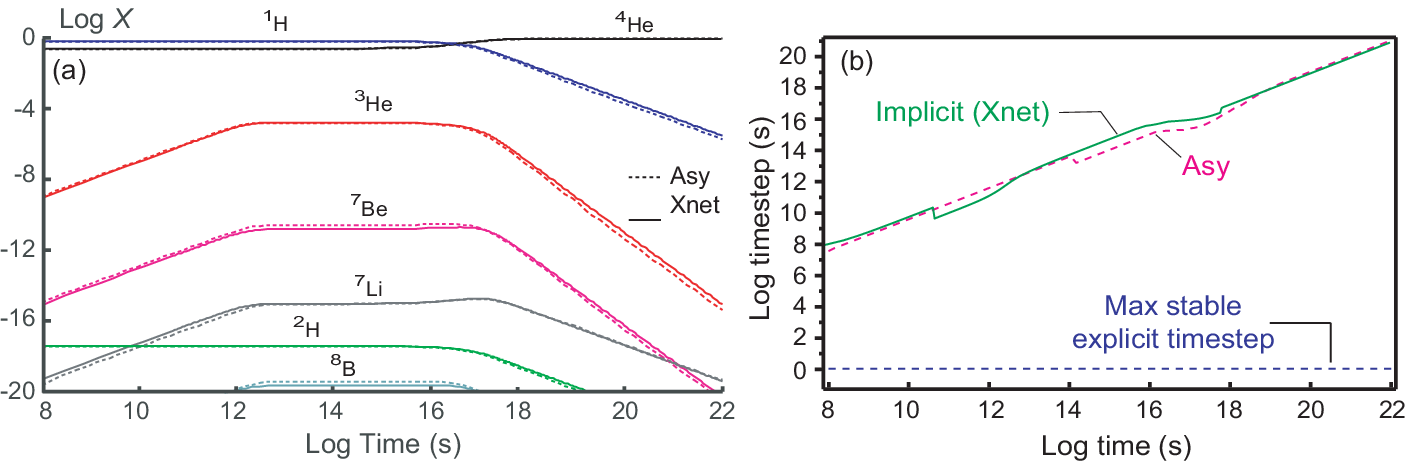}
     {\figdn}
     {\figdn}
     {0.97}
{Integration of the pp-chains under constant temperature and density
characteristic of the core of the present Sun: $T_9 = 0.016$ and a density of
$160 \units{g\,cm}^{-3}$, assuming solar initial abundances. Reaction rates were
taken from the REACLIB library \cite{raus2000}.  (a)~Mass fractions for the
asymptotic method of Eq.\ (\ref{asySimple}) (dotted curves) and for the standard
implicit code Xnet \cite{raphcode} (solid curves). (b)~Integration timesteps for
the asymptotic method (dotted magenta) and the implicit method (solid green).
The expected maximum stable fully-explicit timestep is indicated by the dashed
blue curve.}
we illustrate integration of the pp-chains at a constant temperature and density
characteristic of the core in the present Sun, using the asymptotic method and
the implicit backward-Euler code Xnet \cite{raphcode}. We see that the explicit
asymptotic integration gives results for the mass fractions in rather good
agreement with the implicit code over 20 orders of magnitude, and is generally
taking timesteps $dt \sim 0.1 t$  that are comparable to those for the implicit
code over the entire range of integration. (The asymptotic method required
333 total integration steps versus 176 steps for the implicit code.) 

As shown in \S \ref{sh:large-stiff}, the maximum stable timestep for a standard
explicit integration method may be estimated as the inverse of the fastest rate
contributing to the network. This is illustrated by the dashed blue curve in
\fig{ppChains}(b). At late times the explicit integration timesteps are
$\sim10^{21}$ times larger than the maximum stable timestep for a normal
explicit integration. The calculation illustrated in \fig{ppChains} takes less
than a second on a 3 GHz processor with the explicit asymptotic method (as does
the implicit solver).  In contrast, from \fig{ppChains}(b) we estimate that a
standard explicit method taking the largest stable fully-explicit timestep would
require a time about 1000 times longer than the age of the Universe ($ \sim
10^{21} \units s$ of processor time) to compute the pp-chains to hydrogen
depletion.

\subsection{Simulations under Nova Conditions}
\protect\label{asynovaConditions}

The preceding example entailed an exceedingly stiff but rather small network.
Let us now turn to an example that is highly stiff and involves hundreds of
isotopes. In \fig{nova125D_XplusHydroProfile}(a)%
\singlefig
     {nova125D_XplusHydroProfile}
     {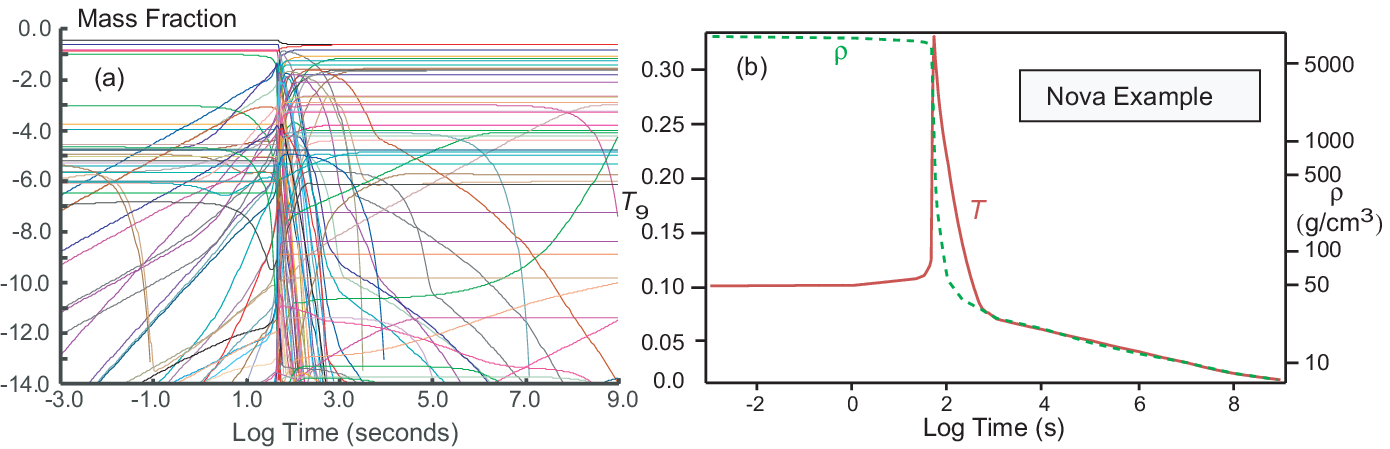}
     {0pt}
     {\figdn}
     {1.0}
{(a)~Mass fractions for a network under nova conditions, corresponding to the
hydrodynamical profile shown in (b). The calculation used the explicit
asymptotic method corresponding to \eq{asySimple} and a network containing 134
isotopes coupled by 1531 reactions, with rates taken from the REACLIB library
\cite{raus2000} and initial abundances enriched in heavy elements
\cite{parete-koon03}.}
we illustrate a calculation using the explicit asymptotic algorithm and a
hydrodynamical profile shown in \fig{nova125D_XplusHydroProfile}(b) that
is characteristic of hot-CNO burning in a nova outburst. The explicit asymptotic
timesteps are displayed in \fig{nova125D_dtPlusfraction}(a).%
\singlefig
     {nova125D_dtPlusfraction}
     {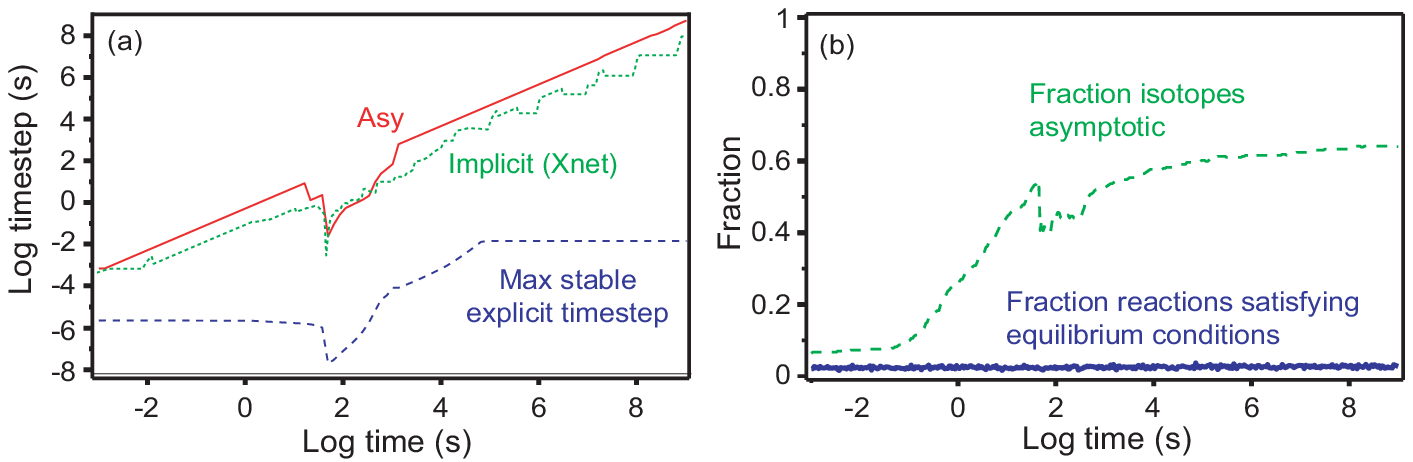}
     {0pt}
     {\figdn}
     {0.95}
{(a)~Timesteps for integration of \fig{nova125D_XplusHydroProfile}. The solid
red curve is from the asymptotic calculation.  The dotted  green curve is from
an implicit integration using the backward-Euler code Xnet \cite{raphcode}.  The
dashed blue curve estimates the largest stable fully explicit timestep as the
inverse of the fastest rate in the system. (b)~Fraction of isotopes that become
asymptotic and fraction of reactions that reach partial equilibrium in the
asymptotic-method calculation.}
In these simulations we see that the  asymptotic network takes stable and
accurate timesteps corresponding to $dt \sim 0.1 t$ over most of the time
integration, except in the region of sharp temperature rise and strong burning,
where $dt \sim (0.01-0.001)t$. Over most of the range of integration after
burning commences, the asymptotic solver timesteps (solid red curve in
\fig{nova125D_dtPlusfraction}(a))  are a million or more times larger than the
maximum stable timestep for a purely explicit method (dashed blue curve in
\fig{nova125D_dtPlusfraction}(a)); at late times this disparity increases and by
the end of the calculation  the asymptotic timesteps are approximately $10^{10}$
times larger than would be possible for a normal explicit integration. 

The generally large explicit asymptotic timesteps over the entire integration
range illustrated in \fig{nova125D_dtPlusfraction}(a) are greater than or equal
to those for a typical implicit code, as may be seen by comparing with the
implicit (backward Euler) calculation timestepping curve shown in dotted green.
In this calculation the implicit method required 1335 integration steps while
the explicit asymptotic calculation required only 935 steps, and furthermore we
expect that the explicit timesteps can be computed more quickly than the
implicit timesteps.  For a network with 134 isotopes,  an optimized explicit
code should calculate a timestep perhaps 7 times faster than typical implicit
codes (Table \ref{tb:explicitSpeedup}). These large explicit timesteps are
possible because during the simulation many isotopes become asymptotic but few
reactions reach partial equilibrium, as illustrated in
\fig{nova125D_dtPlusfraction}(b). Similar results for nova simulations were
obtained with the asymptotic method in Refs.~\cite{feg11a,feg11b} using a
different nova hydrodynamical profile and a different reaction library. We
conclude that the explicit asymptotic method may intrinsically be an order of
magnitude faster than a state-of-the art implicit code for simulations under
nova conditions.

\subsection{Simulations under Tidal Supernova Conditions}
\protect\label{asyTidalConditions}

The mass fractions as a function of time for a thermonuclear supernova event
induced by tidal interaction in a white-dwarf \cite{tidalSupernova} are
illustrated in \fig{tidalAlpha_XplusHydroProfile}(a)%
\singlefig
     {tidalAlpha_XplusHydroProfile}
     {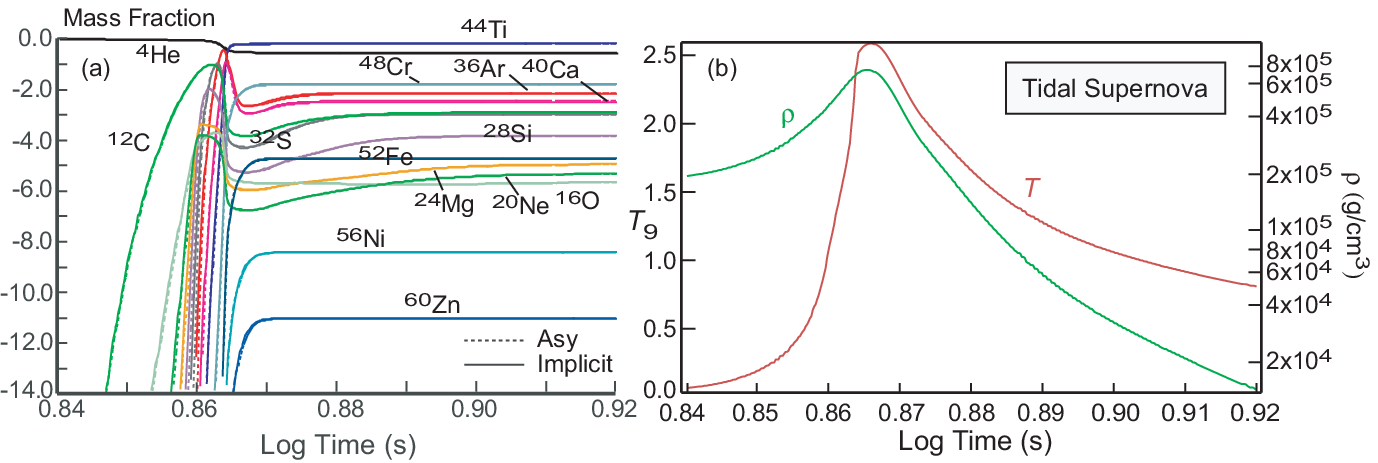}
     {0pt}
     {\figdn}
     {1.0}
{(a)~Mass fractions for an alpha network under tidal supernova conditions with
an initial abundance of pure \isotope{4}{He}. The network contained 16
isotopes coupled by 46 reactions, with rates from REACLIB \cite{raus2000}. 
Dotted curves calculated in asymptotic approximation using \eq{asySimple}. Solid
curves calculated using the fully implicit code Xnet \cite{raphcode}. (b)~The
corresponding hydrodynamical profile \cite{tidalSupernova}.}
% %
% %
for an alpha network and the hydrodynamical profile is illustrated in
\fig{tidalAlpha_XplusHydroProfile}(b). Dotted curves correspond to the explicit
asymptotic calculation and solid curves correspond to a fully-implicit,
backward-Euler calculation using Xnet \cite{raphcode}. The mass fractions for
the two calculations are almost indistinguishable. The timestepping  is compared
for the asymptotic calculation and Xnet in \fig{tidalAlpha_dtPlusFraction}(a).
\singlefig
     {tidalAlpha_dtPlusFraction}
     {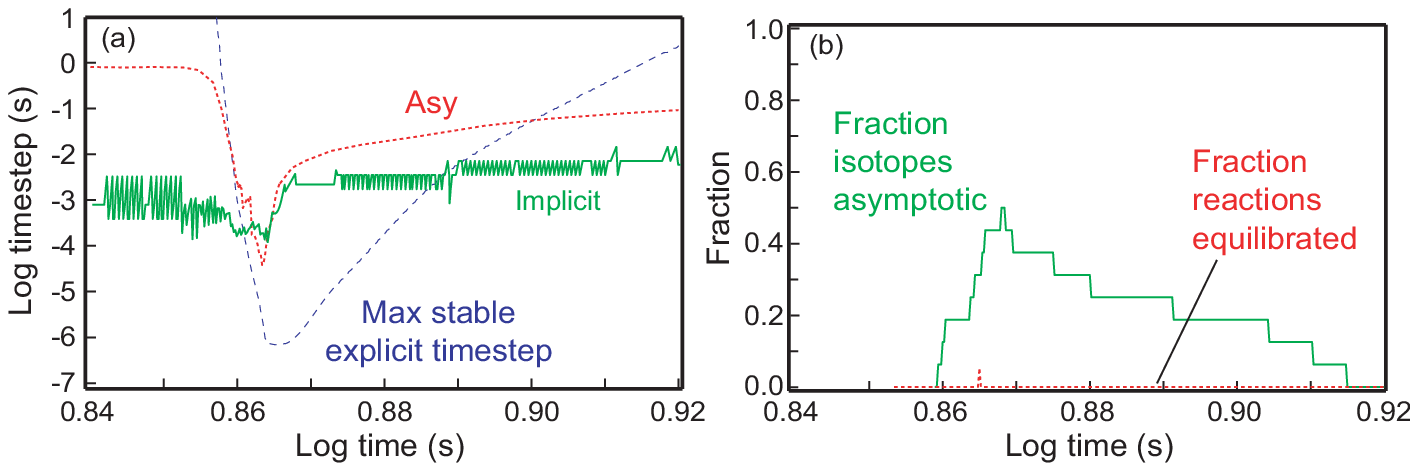}
     {0pt}
     {\figdn}
     {0.95}
{(a)~Asymptotic integration timesteps (dotted red), integration steps for the
implicit code Xnet \cite{raphcode} (solid green) and maximum stable purely
explicit step (dashed blue) for the calculation in
\fig{tidalAlpha_XplusHydroProfile}. (b)~Fraction of isotopes that become
asymptotic and fraction of reactions equilibrated in the network.}
% %
% %
We see that the timestepping for the asymptotic calculation is somewhat better
than for the implicit code (480 total integration steps for the asymptotic
calculation versus 2136 total integration steps for the implicit calculation).
Estimating that a fully-optimized explicit solver can calculate a timestep about
3 times faster than an implicit code like Xnet in a 16-isotope network (Table
\ref{tb:explicitSpeedup}), we surmise that the asymptotic method is capable of
doing the integration for \fig{tidalAlpha_dtPlusFraction} perhaps 10 times
faster than a current implicit code. The relatively good timestepping for the
asymptotic method in this case is because essentially no reactions in the
network come into equilibrium, as illustrated in
\fig{tidalAlpha_dtPlusFraction}(b).  The flat mass fraction curves at late times
in \fig{tidalAlpha_XplusHydroProfile}(a) are not a result of equilibrium but
rather of reaction freezeout caused by the temperature and density dropping
quickly at late times as the system expands
(\fig{tidalAlpha_XplusHydroProfile}(b)).

A calculation for the hydrodynamical profile illustrated in
\fig{tidalAlpha_XplusHydroProfile}(b) but for a 150-isotope network is
illustrated in \fig{tidal150Xdt}.%
\singlefig
     {tidal150Xdt}
     {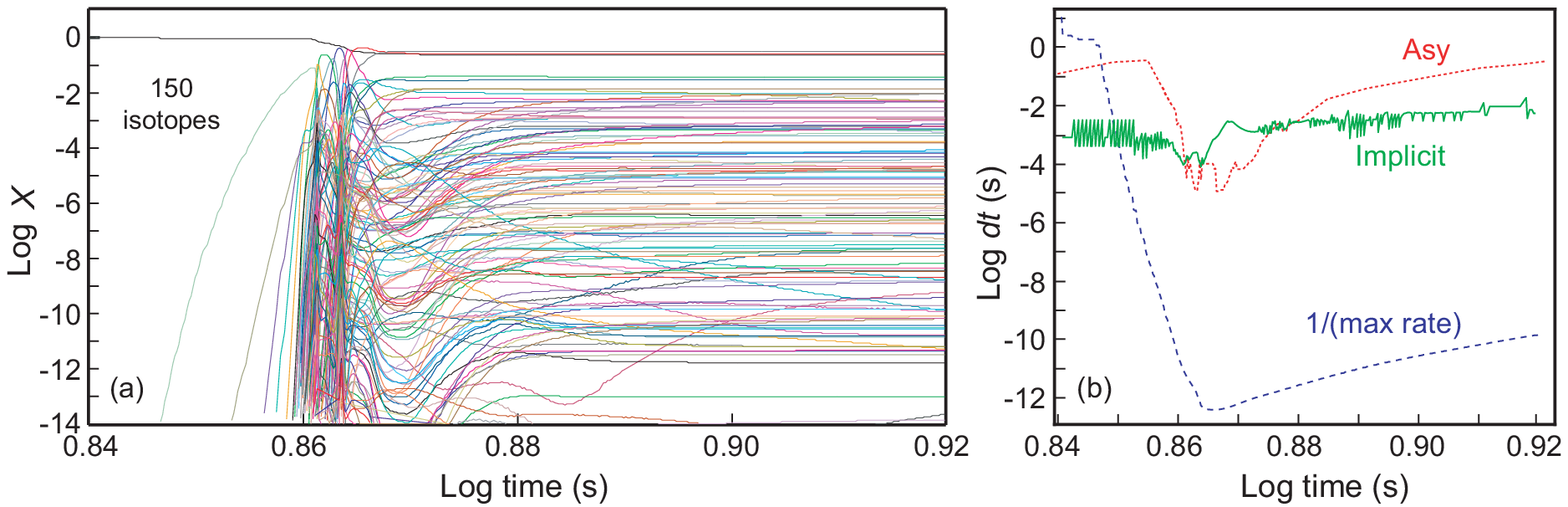}
     {0pt}
     {\figdn}
     {0.72}
{(a)~Mass fractions for a 150-element network with 1563 reaction couplings under
tidal supernova conditions, corresponding to the hydrodynamical profile shown in
\fig{tidalAlpha_XplusHydroProfile}(b). Initially the abundance was pure
\isotope{4}{He}   and rates were taken from REACLIB \cite{raus2000}. The mass
fractions were calculated in asymptotic approximation using \eq{asySimple}.
(b)~The corresponding integration timesteps for the asymptotic method (dashed
red) and the implicit code Xnet \cite{raphcode} (solid green). The maximum
stable explicit timestep is estimated by the dashed blue curve.}
% %
% %
We note that in this example the asymptotic approximation permits timesteps that
are typically 8--9 orders of magnitude larger than the maximum stable explicit
timestep (indicated by the dashed blue line). These timesteps are again
competitive with those of a standard implicit code (shown as the solid green
curve in \fig{tidal150Xdt}(b)). In this case the implicit code took 2425
timesteps to complete the integration while the asymptotic method took 5593
integration steps, but this factor of about two fewer implicit timesteps would
be more than offset by the significantly faster computation of each integration
step expected for an optimized explicit asymptotic code integrating a network
this large. For a 150-isotope network we expect that an optimized explicit code
would be more than 7 times faster computing a timestep than an implicit code
(Table \ref{tb:explicitSpeedup}), so a fully-optimized  asymptotic method is
probably capable of doing the integration in \fig{tidal150Xdt} several times
faster than a current implicit code.

A calculation for the hydrodynamical profile illustrated in
\fig{tidalAlpha_XplusHydroProfile}(b) for a 365-isotope network is
illustrated in \fig{tidal365Xdt}.%
\singlefig
     {tidal365Xdt}
     {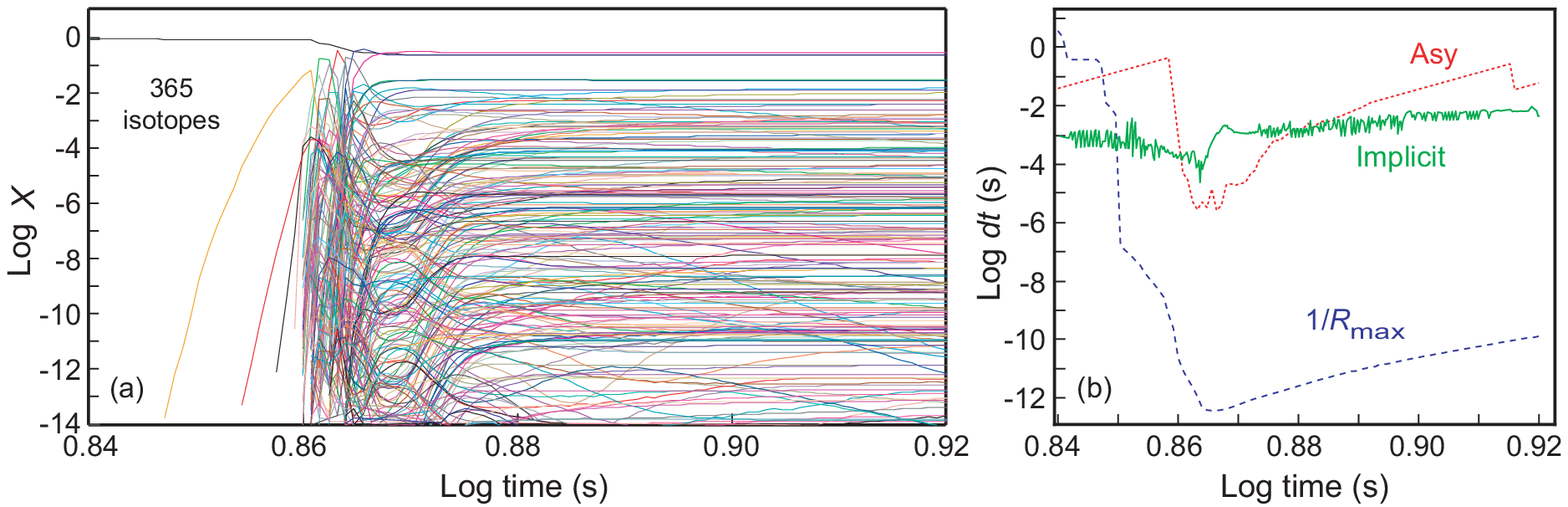}
     {0pt}
     {\figdn}
     {0.72}
{(a)~Mass fractions calculated in asymptotic approximation using \eq{asySimple}
for a 365-element network with 4325 reaction couplings under tidal supernova
conditions, corresponding to the hydrodynamical profile shown in
\fig{tidalAlpha_XplusHydroProfile}(b).  Initially the abundance was pure
\isotope{4}{He} and rates were taken from REACLIB \cite{raus2000}. (b)~The
corresponding integration timesteps for the asymptotic method (dashed red curve)
and the implicit code Xnet \cite{raphcode} (solid green curve). The maximum
stable explicit timestep is indicated by the dashed blue curve.}
% %
% %
We note that in this example the asymptotic approximation permits timesteps that
are as much as $10^{10}$ times larger than the maximum stable explicit timestep
(dotted blue curve). This timestep is again competitive with that of an implicit
calculation (solid green curve in \fig{tidal365Xdt}(b)).  The implicit
calculation required 2707 integration steps, compared with 5778 steps for the
asymptotic calculation. But as noted above, this factor of two advantage of the
implicit calculation should be more than offset by the much more efficient
computation of each timestep in an optimized asymptotic code. For this
365-isotope network we may assume from Table \ref{tb:explicitSpeedup} that the
explicit code can calculate each timestep $\sim 20$  times faster than the
implicit code, so an optimized asymptotic code should be capable of performing
the integration in \fig{tidal365Xdt} perhaps 10 times faster than a
state-of-the-art implicit code.

Similar results for networks under
tidal supernova conditions have been found in
Refs.~\cite{feg11a,feg11b}. Although a different set of reaction network rates
was used in these references, the explicit asymptotic method was again found
to be highly competitive with standard implicit methods for the tidal supernova
problem.

\subsection{\label{sh:scaling} Scaling with Network Size}

  In \fig{scalingNetworkSizeNovaSimulation}%
\singlefig
     {scalingNetworkSizeNovaSimulation}
     {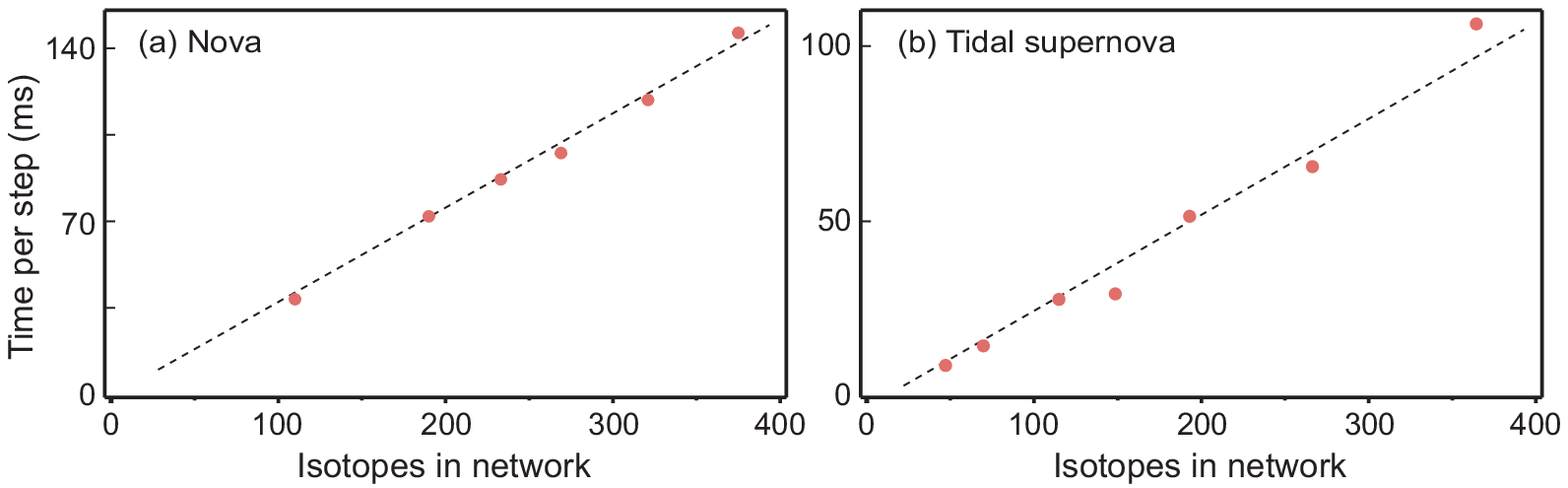}
     {0pt}
     {\figdn}
     {0.87}
{Linear scaling of wall clock time per integration step with number of isotopes
in the network for the explicit asymptotic approximation. (a)~The nova
simulation of \fig{nova125D_XplusHydroProfile}. (b)~The tidal supernova
simulation of
Figs.~\ref{fig:tidalAlpha_XplusHydroProfile}--\ref{fig:tidal365Xdt}. The dashed
lines are drawn only to guide the eye.}
we illustrate scaling of integration time with network size for the nova
simulation of \fig{nova125D_XplusHydroProfile} and the tidal supernova
simulation of
Figs.~\ref{fig:tidalAlpha_XplusHydroProfile}--\ref{fig:tidal365Xdt}. The
behavior is seen to be approximately linear, as expected for an explicit
algorithm since no matrix inversions are required.

\section{\label{sh:asySophisticated} Tests of More Sophisticated Asymptotic
Algorithms}

All results presented to this point have used the simplest asymptotic formula
defined in \eq{asySimple}. This section compares results from Eq.\
(\ref{asySimple}) to those obtained with the more sophisticated formulas
\eqnoeq{asySophia}, and \eqnoeq{asyOB}, for the case of an alpha network with
constant $T_9 = 5$ and $\rho = 1\times 10^8 \units{g\,cm}^{-3}$. In
\fig{XasyCompositeT9_5rho1e8}
 \singlefig
     {XasyCompositeT9_5rho1e8}
     {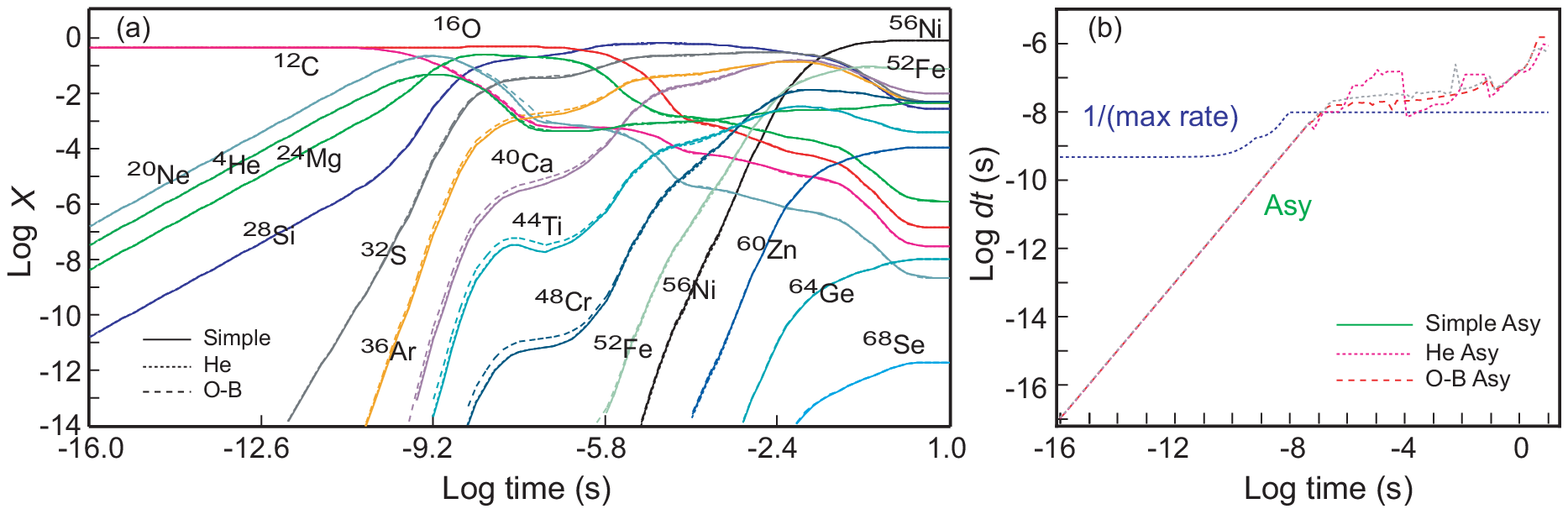}
     {0pt}
     {\figdn}
     {0.72}
{Comparison of mass fractions (a) and timesteps (b) using different asymptotic
approximations for an alpha network with constant $T_9 = 5$ and $\rho = 1\times
10^8 \units{g\,cm}^{-3}$. The network contained 16 isotopes coupled by 46
reactions, with rates taken from REACLIB \cite{raus2000}. Initial abundances
corresponded to equal mass fractions of \isotope{12}{C} and \isotope{16}{O}.
Simple refers to \eq{asySimple}, He to \eq{asySophia}, and O--B to \eq{asyOB}.}
we display a composite of the computed mass fractions and the timestepping for
the asymptotic approximations corresponding to these three formulas. We find
that the more sophisticated asymptotic approximations in Eqs.\
\eqnoeq{asySophia} and \eqnoeq{asyOB}  give results for abundances that are very
similar to those from \eq{asySimple}, but can in some cases give more favorable
timestepping by factors of several. These results are representative of tests on
a variety of networks and we conclude that for typical thermonuclear networks
Eqs.\ \eqnoeq{asySimple}, \eqnoeq{asySophia}, and \eqnoeq{asyOB} yield similar
results, except for possible differences by factors of up to 2--3 in
computational speed.

\section{\label{sh:noncompetive} Non-Competitive Asymptotic Timesteps in the
Approach to Equilibrium}

In previous sections evidence has been presented that, well-removed from
equilibrium, asymptotic methods can provide stable and accurate integration of
the stiffest large networks with timesteps that are comparable to those employed
in standard implicit and semi-implicit stiff solvers. In practice, for
astrophysical thermonuclear networks this means that timesteps are typically
from $0.1$ to $0.001$ of the current time over most of the integration range,
except for brief time periods where very strong fluxes are being produced and
timesteps may need to be shorter to maintain accuracy.  Since explicit methods
can generally compute each timestep substantially faster than for implicit
methods, this suggests that such methods offer a viable alternative to implicit
solvers under those conditions. 

However, the preceding statements are no longer true when substantial numbers of
reaction pairs in the network begin to satisfy microscopic equilibrium
conditions (according to the criteria given in \S\ref{ss:measureEquil}). Then
the typical behavior for asymptotic  approximations is for the timestep to
become constant or only slowly increasing with integration time. (Indeed, we
have already seen an indication of this behavior at late times in
\fig{XasyCompositeT9_5rho1e8}(b)). Figure
\ref{fig:dtHydro_withBand_withNetwork_withEquilFracComposite}%
%
%
%Source:/home/guidry/mwg/programming/flashCases/oneZone/T9_3rho_1e7/approx19/
\putfig
{dtHydro_withBand_withNetwork_withEquilFracComposite}
{0pt}
{\figdn}
{0.65}
{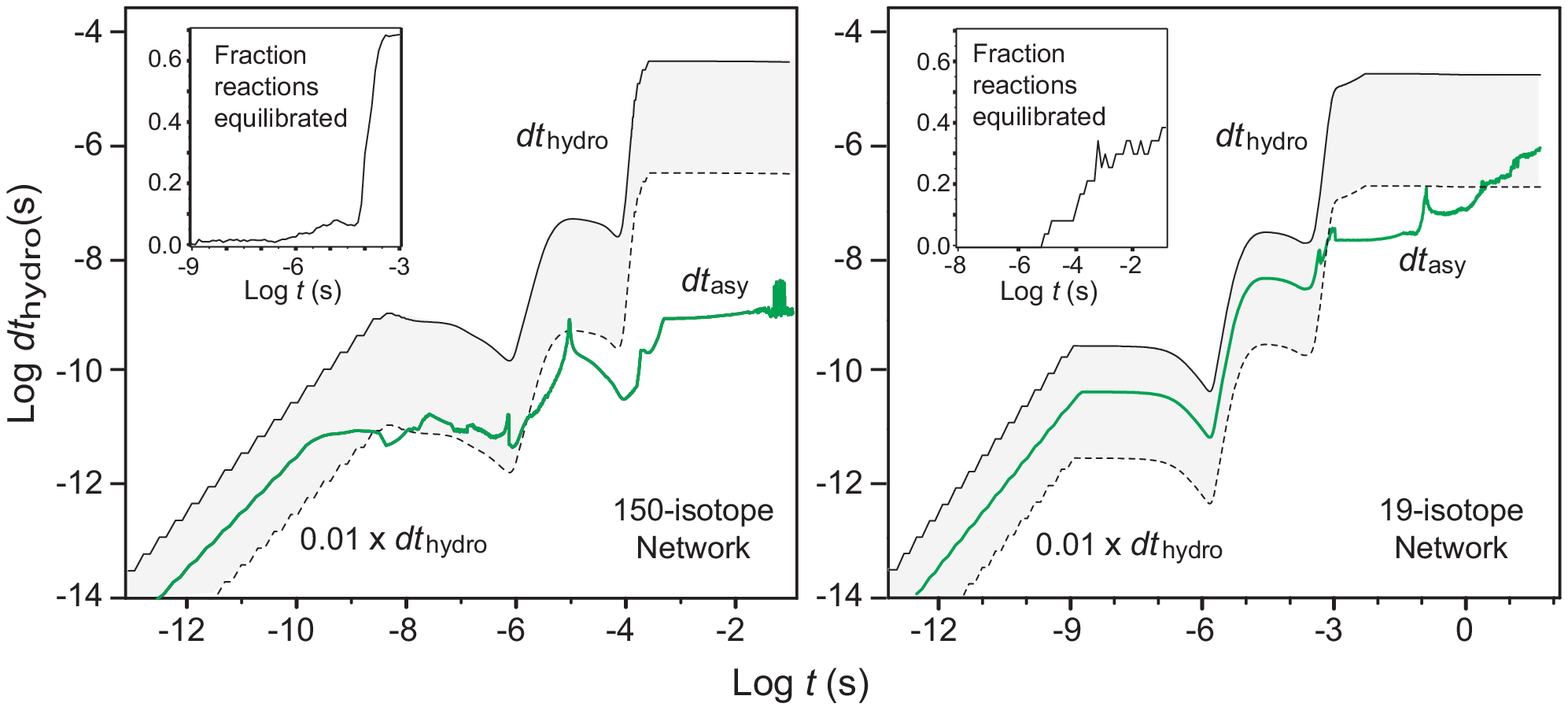} {Network
timesteps (solid green curves) in asymptotic approximation for a single-zone
2-dimensional hydrodynamics simulation using the Flash \cite{fry00}
hydrodynamics code. Left: 150-isotope network. Right: 19-isotope network. Insets
show the fraction of reactions that reach equilibration as a function of time.
The shaded regions indicate the target network timestepping required for the
network to keep up with the hydrodynamics, as discussed further in conjunction
with Fig.~\ref{fig:dtHydro}.}
% %
% %
illustrates for two different networks coupled to a single-zone hydrodynamical
simulation. We see that the 19-isotope network is able only marginally to keep
up with the hydrodynamical timesteps, while the 150-isotope network lags orders
of magnitude behind for hydrodynamical integration times later than about
$10^{-4}-10^{-5}$ s. 

The reason for this loss of timestepping efficiency for the
asymptotic method is the approach to equilibrium at late times; this is
documented in the inset plots for
\fig{dtHydro_withBand_withNetwork_withEquilFracComposite}, which show the
fraction of reactions in the respective networks that satisfy equilibrium
conditions.  The reason that the 19-isotope network is able to keep up much
better than the 150-isotope network is also clear from the inset plots: the
19-isotope network slowly approaches 30--40\% equilibration in this region, but
the 150-isotope network, with much faster reactions because it includes fast
proton and neutron reactions not found in the 19-isotope network, quickly
reaches 70\% equilibration. Since in most applications for extremely stiff
astrophysical networks the physical phenomena require integration over many
decades of time, this lag of the asymptotic timestepping as equilibrium is
approached is disastrous for such approximations and they quickly lose out to
implicit methods, which can continue to take large integration steps even
nearing equilibrium (though they are inefficient at computing each timestep).

Another example of the failure of the asymptotic approximation to generate
competitive timesteps is illustrated in \fig{viktorExtended2AlphaComposite}.%
\putfig
{viktorExtended2AlphaComposite}
{0pt}
{\figdn}
{0.95}
{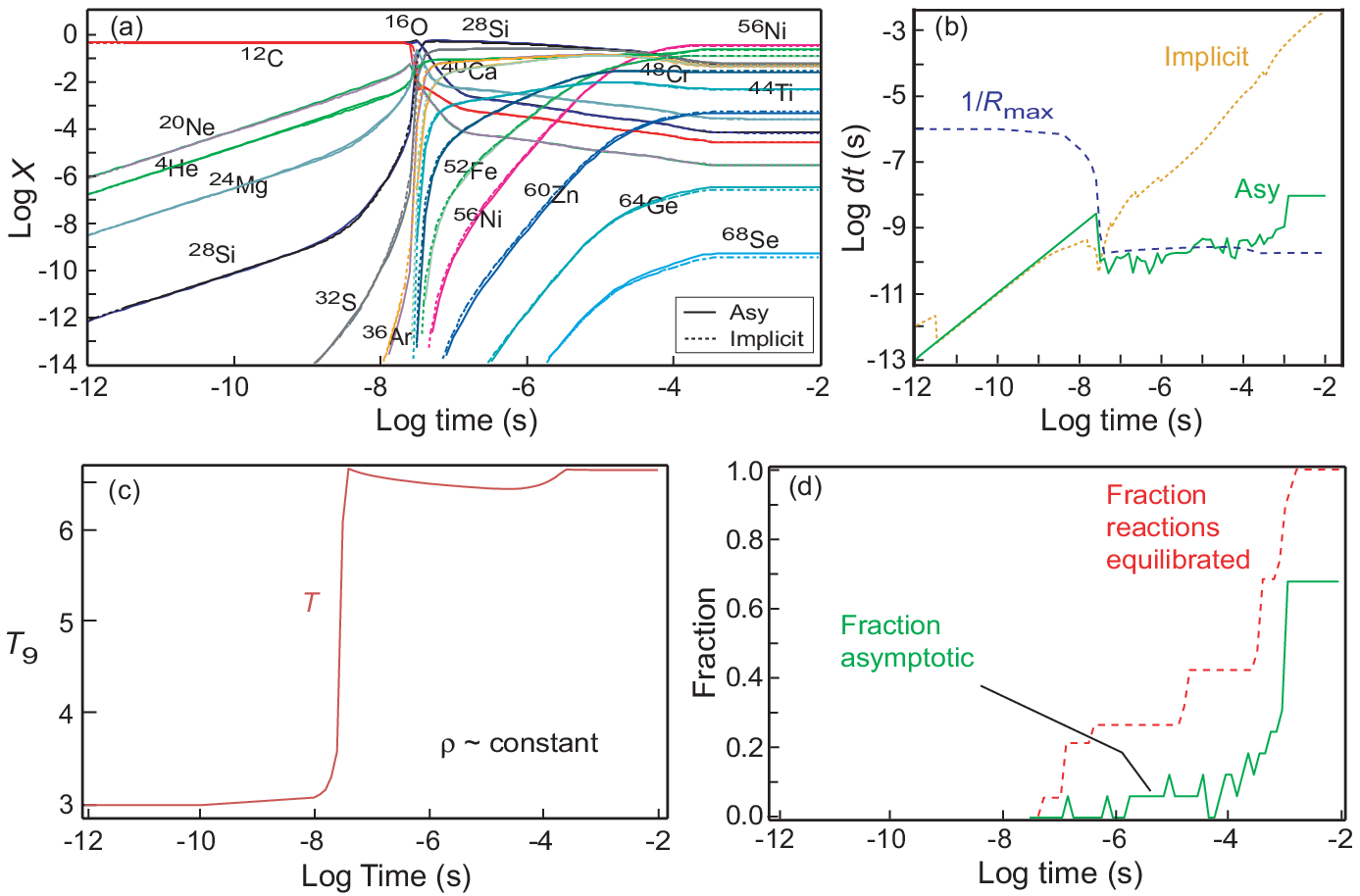}
{Asymptotic calculation with an alpha network for a hydrodynamic profile
characteristic of a single isolated zone in a Type Ia supernova explosion.
(a)~Mass fractions in asymptotic approximation (solid) and with the implicit
code Xnet \cite{raphcode} (dotted). (b)~Integration timesteps. The asymptotic
timesteps are illustrated by the solid green curve and the Xnet implicit
timesteps by the dotted orange curve. The dashed blue curve labeled
$1/R\tsub{max}$ indicates approximately the maximum stable timestep for a purely
explicit method (corresponding to the inverse of the fastest rate in the
network). (c)~The hydrodynamic
profile. (d)~Fraction of isotopes that become asymptotic and the fraction of
network reactions that become equilibrated in the course of the explosion. }
% %
% %
The asymptotic mass fractions calculated in
\fig{viktorExtended2AlphaComposite}(a) are quite accurate in comparison with the
results from standard implicit codes, but the timestepping illustrated in
\fig{viktorExtended2AlphaComposite}(b) is not competitive  with implicit methods
at late times.  For $\log t \sim -2$, the asymptotic integration timesteps are
only of order $10^{-8}$ seconds, whereas the implicit code is taking timesteps
larger than $10^{-3}$ seconds at this point.

The reason for this failure of the asymptotic approximation can be seen clearly
in \fig{viktorExtended2AlphaComposite}(b) and (d). In the region of rapid
temperature rise illustrated in \fig{viktorExtended2AlphaComposite}(c), the
rates in the network increase dramatically and this causes the maximum stable
explicit timestep to become much smaller, as illustrated in
\fig{viktorExtended2AlphaComposite}(b) by the dashed blue curve. The steadily
increasing explicit timestep (solid green curve) intersects this maximum stable
explicit timestep curve near $\log t = -8$.  But from
\fig{viktorExtended2AlphaComposite}(d) we see that essentially no isotopes in
the network satisfy the asymptotic condition at this point.  Thus the
integration is forced to use a standard explicit method and the timestep must
(to maintain stability) follow the decreasing dashed blue curve until around
$\log t = -4$, when significant numbers of isotopes finally begin to satisfy the
asymptotic condition and the explicit asymptotic algorithm is able begin to take
timesteps larger than the explicit limit.  However, at this point the asymptotic
timesteps are already orders of magnitude smaller than those an implicit method
would use, and the asymptotic method is able to increase the timestep by only
about two orders of magnitude before the system reaches equilibrium.  Thus, for
the entire time range from $\log t \sim -8$ until $\log t \sim -2$ the explicit
asymptotic method computes the network accurately but its timesteps lag many
orders of magnitude behind those from standard implicit methods.

As we shall explain in considerable detail in the third paper of this series
\cite{guidPE}, the reason for the loss of efficiency for asymptotic methods as
equilibrium is approached is that the asymptotic approximation removes a large
amount of stiffness associated with macroscopic equilibration, but near
(microscopic) equilibrium a fundamentally new source of stiffness begins to play
a role and it is not generally removed by the asymptotic approximation. Indeed,
we see clearly from \fig{viktorExtended2AlphaComposite}(d) that the fraction of
reactions in the network that satisfy equilibrium conditions increases rapidly
beginning at $\log t \sim -7$ and reaches unity by $\log t \sim -3$.   In this
third paper we shall describe a new implementation of {\em partial equilibrium
methods} that can be used in conjunction with asymptotic methods to turn
equilibrium from a liability into an asset and increase the explicit
timestepping by orders of magnitude in the approach to equilibrium. In that
paper we will give examples suggesting that these methods are capable of giving
timestepping competitive with that of implicit methods across the entire range
of interesting physical integration times for a variety of extremely stiff
reaction networks.

\section{Summary and Conclusions}

Explicit numerical integration can compute a timestep faster then implicit
methods, and the time to compute a network explicitly scales linearly and
therefore more favorably with network size than for implicit codes.
Nevertheless, previous discussions of numerical integration for very stiff
systems have concluded rather uniformly that explicit methods are not
competitive with implicit methods for stiff networks because they are unable to
take large enough stable timesteps. To quote {\em Numerical Recipes}
\cite{press92}, ``For stiff problems we {\em must} use an implicit method if we
want to avoid having tiny stepsizes.''  

Improvements in explicit methods based on using asymptotic and steady-state
limiting solutions to remove stiffness from the network have had some success
for systems of moderate stiffness such as various chemical kinetics problems.
However, it has been concluded in the previous literature that such methods are
not competitive, failing even to give correct results, with timesteps that are
far too short to be useful even if they gave correct results, for extremely
stiff networks such as those encountered commonly in astrophysical thermonuclear
networks \cite{oran05,mott99}. This paper has presented evidence strongly
challenging all of these conclusions. 

% See page 742 in Numerical Recipes (Fortran, 1992) for above quote

We have cleanly identified three fundamentally different sources of stiffness
in large networks, only the first of which is commonly emphasized in the
literature:
\begin{enumerate}
 \item 
Situations where small populations can become negative if the explicit
timestep is too large, with the propagation of this anomalous negative
population leading to exponentially growing terms that destabilize the network.

\item
Situations where the right sides of the differential equations expressed as $dY
= F = \fplus{} - \fminus{}$ approach a constant derived from the difference of
two large numbers (the total flux in $\fplus{}$ and total flux out $\fminus{}$),
and numerical errors in taking this difference destabilize the network if the
timestep is too large. 

\item
Situations where on the right sides of the differential equations expressed in
the form of Eq.~(\ref{equilDecomposition}) the net flux in specific
forward-reverse reaction pairs $(f^+_i - f^-_i)$ tends to zero as the system
approaches equilibrium, leading to large errors if the timestep is too large
because the net flux is derived from the difference of two large numbers and
the timescale equilibrating the populations is short compared with the desired
numerical timestep.  

\end{enumerate}
We have shown that these distinctions are important because different sources of
stiffness require different approximations for their removal in an
algebraically-stabilized explicit integration. 

Using the extremely stiff systems characteristic of astrophysical thermonuclear
networks as a stringent test, we have shown that asymptotic methods are very
successful at removing the first two types of stiffness, and give correct
results, even for the stiffest of thermonuclear networks, provided that adequate
attention is paid to conservation of probability in the network.  Furthermore,
we have shown various examples of stable and accurate timestepping with these
methods in extremely stiff systems that are competitive with that of standard
implicit codes, demonstrating in some simple but physically-important networks
timesteps that are as much as 20 orders of magnitude larger than the maximum
timestep that would be stable in a standard explicit method. 

Asymptotic methods are adept at removing the first two types of stiffness listed
above, permitting explicit numerical timesteps that are competitive with
implicit methods even in the stiffest networks.  However,  we have also shown
that such methods give correct results but fail to exhibit competitive
timestepping when the system approaches microscopic equilibrium and the third
type of stiffness instability begins to dominate. In a following paper
\cite{guidPE}, we shall provide evidence for competitive timestepping, even in
the approach to equilibrium, if the explicit asymptotic method is supplemented
by partial equilibrium approximations designed specifically to deal with the
third type of stiffness instability. 

Taken together, this paper and the following ones on quasi-steady-state methods
\cite{guidQSS} and partial equilibrium methods \cite{guidPE} present compelling
evidence that algebraically-stabilized explicit integration methods are capable
of timesteps competitive with implicit integration methods for a variety of
highly-stiff reaction networks. Since explicit methods can execute a timestep
faster than an implicit method in a large network, our results suggest that
algebraically-stabilized explicit algorithms may be capable of performing as
well as, or even substantially outperforming, implicit integration in a variety
of moderate to extremely stiff applications. Because of the highly-favorable
linear scaling for explicit methods, this fundamentally new view of the efficacy
of explicit integration for stiff equations may be particularly important for
applications in any field where it is imperative that more realistic---and
therefore much larger---networks be used in complex physical simulations.

\begin{ack}
We thank Tony Mezzacappa for useful discussions, Austin Harris for help with
some of the calculations, Eric Lingerfelt for programming assistance, and
Christian Cardall for a careful reading of the manuscript. Research was
sponsored by the  Office of Nuclear Physics, U.S. Department of Energy.

\end{ack}

\clearpage

\bibliographystyle{unsrt}

\end{document}